\documentclass[aps,prl,twocolumn,showpacs,superscriptaddress,floatfix,longbibliography]{revtex4-1}
\usepackage{graphics}
\usepackage[USenglish]{babel}
\usepackage{graphicx}
\usepackage{amssymb}
\usepackage{amsmath}
\usepackage{xcolor}
\usepackage{natbib}

\newcommand{\bs}[1]{\boldsymbol{#1}}                      
   
\begin{document}
	\preprint{APS}
	
	\title{
Oscillating, non-progressing flows induce directed cell motion
	}
	
	\author{Winfried Schmidt}
	\affiliation{Theoretische Physik, Universit\"at Bayreuth, 95440 Bayreuth, Germany}
	\affiliation{Laboratoire Interdisciplinaire de Physique, Universit\'e Grenoble Alpes and CNRS, F-38000 Grenoble, France}
	
	\author{Andre F\"ortsch}
	\affiliation{Theoretische Physik, Universit\"at Bayreuth, 95440 Bayreuth, Germany}
	
	\author{Matthias Laumann}
	\affiliation{Theoretische Physik, Universit\"at Bayreuth, 95440 Bayreuth, Germany}

	\author{Walter Zimmermann}
	\email[]{Corresponding author: walter.zimmermann@uni-bayreuth.de}
	\affiliation{Theoretische Physik, Universit\"at Bayreuth, 95440 Bayreuth, Germany}

	\date{\today}
	\begin{abstract}
We present a deformation-dependent propulsion phenomenon for soft particles such as cells in microchannels. It is based on a broken time reversal symmetry generated by a fast forward and slow backward motion of a fluid which does not progress on average. In both sections, soft particles deform differently and thus progress relatively to the liquid. We demonstrate this by using Lattice-Boltzmann simulations of ubiquitous red blood cells in microchannels, as well as simulations for capsules and minimal soft tissue models in unbounded Poiseuille flows. The propulsion of the soft particles depends besides the oscillation asymmetry on their size, deformation type and  elasticity. 
This is also demonstrated by analytical calculations for a minimal model.
Our findings may stimulate a rethinking
of particle sorting methods. For example, healthy and malignant cells  often differ in their elasticity. With the proposed method, several cell types with different deformability can be separated simultaneously without labeling or obstacles in a microfluidic device.
	\end{abstract}
	
	\maketitle
	

The massive growth of the field of microfluidics is due to a number of recent advances, including methods for focusing and sorting microparticles, such as healthy and cancerous cells \cite{DiCarlo:2017,Nguyen:2019,LimCT:2010.2,Karimi:2013.1,Sajeesh:2014.1,LopezG:2015.1,Kumar_S:2015.1,ParkJY:2016.1,DiCarlo:2019.1,Nasiri:2020.1,Toonder:2020.1,LiuZ:2020.1}. In particular,  label-free hydrodynamic separation methods are finding increasing applications due to their robustness. Here, we report on a microfluidic transport process based on a symmetry breaking and appropriate for separation  of  particles of different elasticity, such as healthy from diseased cells.
  
The change in deformability 
of individual cells has proven to be a useful indicator for the detection of diseases such as cancer \cite{SURESH2007413,GUCK20053689}, blood diseases (sickle cell anemia) \cite{SaadS:2003.1}, inflammation \cite{MacNee:2002.1}, malaria \cite{Dondrop:2000.1} or diabetes \cite{McMillan:1978.1}. In particular, the stiffness of individual cancer cells is drastically reduced compared to normal tissue of the same origin. Furthermore, decreasing single cell stiffness correlates with increasing invasiveness or metastatic potential.  From this perspective, there is a great need for methods to safely separate cells of different stiffness.
  
The widely used microfluidic particle separation methods such as filtration \cite{Debnath:2018.1}, inertial microfluidics, including curved microchannels \cite{LiWeihua:2016.1,DiCarlo:2019.1} or deterministic lateral displacement (DLD) \cite{Austin:2004.1,Austin:2006.2,Bridle:2014.1,AustinGomp:2020.1} are successful methods especially for separating solid particles of different sizes.  For these classical separation techniques the deformability of particles is an additional degree of freedom  and ongoing research is addressing the associated effects \cite{Gompper:2019.1,AustinGomp:2020.1}.

In contrast, several transport phenomena in low Reynolds number microfluidics only occur for soft particles. This is the lift force of vesicles and capsules in linear shear flows and Poiseuille flows near symmetry breaking walls \cite{Secomb:2017.1,Misbah:1999.1,Seifert:1999.1,Viallat:2002.1}.
In Poiseuille flows, the shear rate changes across deformable particles and breaks their forward-backward symmetry, so that droplets \cite{Leal:1980.1,Chakraborty:2015.1}, bubbles and capsules
\cite{Kaoui:2008.1,Misbah:2008.1,Bagchi:2008.1}   even in unconfined Poiseuille flows  exhibit cross-stream migration (CSM) toward the center of a parabolic flow profile. The CSM direction is reversed by a sufficiently strong viscosity contrast between the interior of  cells and the surrounding fluid \cite{Farutin:2014.1} or in vertical channels for non-buoyant capsules by gravitational effects \cite{Laumann:2017.2}.  For modulated  microchannels, a secondary attractor occurs aside from the channel center for soft particles such as  red blood cells (RBCs) and capsules \cite{Laumann:2019.1}. 
These examples for CSM are however not approprate to separate particles with respect to a continuous variation of deformability.

Time-periodic flows in microchannels attract increasingly  attention  \cite{Sauret:2020.1,Daibri:2020.1,JohnT:2021.1,Austin:2009.1,McFaul:2012.1,ParkE:2016.1,Kanso:2016.1,Laumann:2017.1,Ishikawa:2018.1,Mutlu7682,Laumann:2019.2}. 
Among others, they are utilized in combination with arrays of asymmetric posts in microchannels to realize deterministic ratchets  \cite{Austin:2009.1} or to reduce clogging in DLD arrays by particles \cite{McFaul:2012.1,ParkE:2016.1}.

We complement time-periodic flows with a crucial symmetry breaking. This is achieved by different forward and backward velocities of a fluid in the microchannel that does not move on average. We study four types of soft particles that are deformed and entrained differently during the two sections of the low Reynolds number flow.  As a consequence, deformable particles are propelled relative to the fluid. This net transport increases with asymmetry of the flow oscillation, particle deformability and size. Thus, our propulsion mechanism is suitable for 
the development of 
promising deformation-sensitive cell sorting methods.


  \begin{figure}[h]
	\begin{center}		
		\includegraphics[width=\columnwidth]{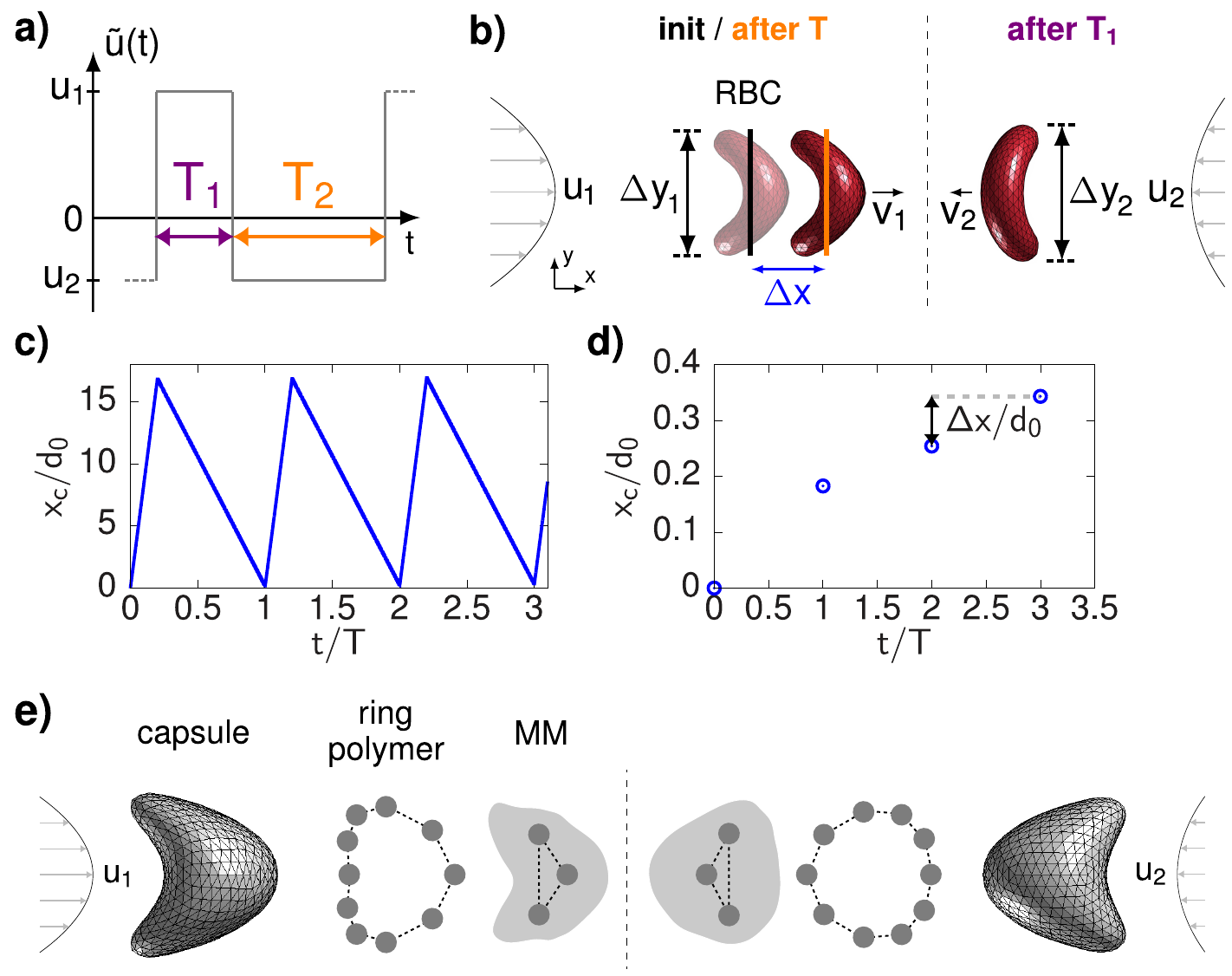}
	\end{center}
	\vspace{-0.4cm}
	\caption{
		a): Time dependence of 
		$\tilde u(t)$ with time intervals  $T_1$ (forward flow, purple), $T_2$ (backward flow, orange)
		and	asymmetry $A=T_2/T_1=2$.
		b): Different RBC shapes during the faster forward and slower backward flow  with $u_1>  |u_2|$, different lateral sizes $\Delta y_1< \Delta y_2$ and velocities $v_1 > |v_2|$. The resulting net progress $\Delta x$ (blue) per flow period $T=T_1+T_2$ is the difference between the RBC's initial (shaded snapshot, black bar) and its final position (bold snapshot, orange bar). 
		c): RBC position $x_\text{c}(t)$ along the channel axis in units of its undeformed diameter $d_0$ obtained by a LBM simulation as function of time for three asymmetric flow cycles with $A=4$. d) Local minima of $x_\text{c}(t)$ with propulsion step $\Delta x$.
		e): Simulation snapshots for capsule, ring polymer and minimal model during forward (left) and backward flow section (right). 	
	}
	\label{fig_sketch_RBC_two_amplitudes}
\end{figure}

The pulsating, undisturbed flow between the plane channel boundaries at $y=\pm w$ is given by
\begin{equation}\label{eq_background_flow}
\bs{u} \left( \bs{r}, t \right) = \tilde{u} (t) \left( 1- \frac{y^2}{w^2} \right) \hat{e}_x\,,
\end{equation}	%
with the unit vector $\hat{e}_x$ in \hbox{$x$-direction}.  We consider  a rectangular time-dependence
\begin{equation}
\label{eq_piecewise_constant_flow_amplitude}
\tilde{u} (t) =
\begin{cases}
u_1 > 0 \quad &\text{for} \quad t \in [0, T_1 [ \\
u_2 < 0 \quad &\text{for} \quad t \in [T_1, T[
\end{cases}
\,~~,
\end{equation}
repeating periodically with flow period $T=T_1+T_2$.
For non-progressing flows the velocities $u_{1,2}$ are related to the time intervals $T_{1,2}$ via
\begin{align}
u_1 T_1 + u_2 T_2 = 0\,~ \Leftrightarrow~  A=\frac{T_2}{T_1}= -\frac{u_1}{u_2}\,,
\end{align}
with the oscillation asymmetry $A$.  
The time-dependent flow amplitude
$\tilde u(t)$ is sketched in Fig.\,\ref{fig_sketch_RBC_two_amplitudes} a) for $A=2$.


We consider four types of soft particles. The minimal model (MM), the 2D ring polymer, and the 3D capsule are represented by $N$ beads with radius $a$. The MM consists of three beads connected by three Hookean springs with spring constant $k$ and equilibrium lengths $b$ and $2b$. It is perpendicularly oriented to $ \hat{e}_x$.
The ring polymer model is represented by a closed bead-spring chain with ten beads placed in the  $x$-$y$- (shear) plane. They are connected by Hookean springs and a bending potential \cite{PhysRevE.51.2658} along the chain provides a ring shape in the quiescent liquid.
The capsule is built of a triangular mesh where the beads are situated at the $N=642$ nodes. We use the Neo-Hookean law \cite{BarthesBiesel:2016.1} for the strain energy with surface shear-elastic modulus $\kappa_\text{S}$, and a bending potential \cite{Gompper:1996.1} with bending rigidity $\kappa_\text{B}$. A volume potential \cite{KruegerT:2013.1} with volume modulus $\kappa_\text{V}$ penalizes deviations from the reference volume $V_0 = 4 \pi r_0^3 / 3$ of the capsule's spherical initial shape with radius $r_0$.
The center of each particle is at $\bs{r}_\text{c} = \sum_{i=1}^{N} \bs{r}_i / N$ with bead positions $\bs{r}_i = (x_i,y_i,z_i)$. The Stokesean dynamics of the beads is described by
\begin{equation}\label{eq_motion_stokesian}
\dot{\bs{r}}_i = {\bs u} ({\bs r}_i, t) + \sum \limits_{j=1}^N {\bs H}_{ij} \cdot {\bs F}_j.
\end{equation}
The forces acting on the beads are ${\bs F}_j = - \nabla_j E({\bs r})$ with the total particle-specific potential $E({\bs r})$ as described above and  ${\bs H}_{ij}$  is the mobility matrix
\cite{Dhont:96,Rotne:1969.1,Wajnryb:2013.1}. Particle-wall interactions are neglected.


The RBC consists of a mesh with the same refinement as the capsule, but has a biconcave initial shape. The strain energy is modeled by the Skalak law \cite{Skalak:1973.1,KruegerT:2013.1} with strain modulus $\kappa_\text{S}$ and area dilation modulus $\kappa_\alpha$. A bending potential according to Ref.\ \cite{MeyerM:2003.1} is used and volume conservation is implemented as for the capsule. The RBC's initial orientation is chosen so that the rotational symmetry axis of its initial shape is parallel to the flow direction.
For simulations of the RBC we use the Lattice-Boltzmann method (LBM) with the Bhatnagar-Gross-Krook collision operator and the immersed-boundary method \cite{Peskin:2002.1,KruegerT:2011.1,KruegerT:2016}.
 The LBM inherently accounts for hydrodynamic interactions (HI) of particles with the channel boundaries. 
For more information on simulation methods and particle models, see Ref.\ \cite{supplement}.
%
If not stated  otherwise, the simulation parameters in Ref.~\cite{all_parameters} are used.
%


Fig.\,\ref{fig_sketch_RBC_two_amplitudes} b) shows simulation snapshots of the RBC with typical croissant-like shapes \cite{Guckenberger:2018.1} during the faster forward and slower backward flow. The two different deformations of the RBC are characterized by the vertical extensions $\Delta y_{1,2}$. This leads to different particle velocities along the channel axis, $v_1 > 0$ and $v_2 < 0$, with
$v_1 \not = |v_2|$. Despite vanishing mean flow, the RBC is propelled after each flow cycle by a step $\Delta x > 0$.
The RBC's $x$-position is shown 
as a function of time in a pulsatile flow with $A=4$ in Fig.\,\ref{fig_sketch_RBC_two_amplitudes} c).
The particle follows
the flow alternately in positive and negative $x$-direction,
with the non-zero propulsive step
relative to the liquid shown in Fig.\,\ref{fig_sketch_RBC_two_amplitudes} d).
Fig.\,\ref{fig_sketch_RBC_two_amplitudes} e) shows the corresponding snapshots for capsule, ring polymer and MM. 

The flow-induced deformation of a particle depends essentially on the curvature of the flow profile 
in Eq.~(\ref{eq_background_flow}), the particle size, the liquid viscosity and the particle elasticity. 
These dependencies can be summarized by the dimensionless capillary number
\begin{align}
\label{eq_Capillary}
 C=\frac{2 |\tilde u| r_0 t_R}{w^2}\,,
 \end{align}
with $C=C_{1,2}$ in the two flow sections $T_{1,2}$.
Here, $2 |\tilde u|/w^2 = |\partial_y^2 u_x(y,t)|$ is the curvature of a plane Poiseuille flow and $2r_0=d_0$ the initial diameter of the particle. For the MM, $d_0=2b$ holds and for the RBC $d_0$ refers to the large diameter of its initial biconcave shape. The particle's relaxation time is given by $t_R=\zeta/k$ for the MM and the ring polymer and by $t_R=\eta r_0 / \kappa_\text{S}$ for the capsule and the RBC, where $\zeta=6 \pi \eta a$ is the Stokes friction and $\eta$  the fluid viscosity.
The particle deformation depends also on 
hydrodynamic particle-wall interactions, but
they  provide only corrections to the leading order bulk effect as shown below for the RBC.


RBC and MM in Fig.\,\ref{fig_sketch_RBC_two_amplitudes} adapt a curved shape that is induced by the $y$-dependence of $u_x(y,t)$ and the associated frictional forces acting on the particle.  
Therefore, as the capillary number $C$ increases, RBC and MM become more curved, which in turn leads to a decrease in the vertical extent, i.e. $\Delta y_1< \Delta y_2< d_0$. The particle velocity $v$ results from an averaging of the local incident flow velocity over the particle surface.  Accordingly, an increasing $C$ results in a decreasing averaging length $\Delta y$ and therefore in an increasing ratio  $v/\tilde u$. In other words, the particle's lag behind with respect to the incident flow at its center is smaller in the forward than in the backward section. Thus, $ v_1/\tilde u_1> |v_2/\tilde u_2|$ and the propulsion step $\Delta x=v_1 T_1 + v_2 T_2$ is positive.

In  the limit of small deformations, the quantities $\Delta x$ and $\Delta y$ can
be determined analytically in terms of the particle properties for the MM. 
The velocities $v_{1,2}$ determine the propulsion velocity according to
\begin{align}
 \label{eq_analytic_actuation_velocity_def}
	v_\text{p} = \frac{\Delta x}{T}  = \frac{v_1 + A v_2}{A+1}~.
	\end{align}
With the explicit  expressions for $v_1$ and $v_2$ (see supplement \cite{supplement}) the propulsion velocity is given by
\begin{align}\label{eq_analytic_actuation_velocity_final}
	v_\text{p} \approx \frac{b}{3 t_R} ~ B_1  \frac{W(A) A^{\frac{1}{3}} - W(1)}{A+1}
\end{align}
with $W(X)=\sqrt{1 + 8X^{\frac{2}{3}}    \left(4X^{\frac{2}{3}} +B_1\right)^{-1}}$ 
and $B_{1,2}=\left(2 C^2_{1,2}\right)^{1/3}$.
The vertical extension $\Delta y$ of the MM decreases with
increasing $C$ at leading order in the following manner:
\begin{align}\label{eq_lateral_size_analytical}
\Delta y_{1,2} \approx 2 b \left[ 1-\frac{B_{1,2}}{12}+  \mathcal{O}(B_{1,2}^2) \right]\,.
\end{align}
For a symmetric flow pulsation one has $\tilde u_1=-\tilde u_2$ and thus $C_1=C_2$.
According to Eq.~(\ref{eq_lateral_size_analytical}), the flow-induced deformation in the two flow sections is identical with $\Delta y_1=\Delta y_2$. The shape is only mirrored after the change between forward and backward flow. I.e. $v_\text{p}$ vanishes for $A=1$ according to  Eq.\,(\ref{eq_analytic_actuation_velocity_final}).
For $A\not =1$,this mirror symmetry is broken, as shown in Fig.\,\ref{fig_sketch_RBC_two_amplitudes}e)
and $v_\text{p}$ becomes finite. As the asymmetry $A>1$ increases, the difference in vertical extensions, $\Delta y_2-\Delta y_1$, and $v_\text{p}$ also increase. 
Moreover, the propulsive step increases with the capillary number $C_1$ and the particle size given by $b$, but decreases with $t_R$. If one changes the sign of $\tilde u \to -\tilde u$, the propulsive direction (sign of $\Delta x$) also changes.

	
	\begin{figure}[htb]
		\begin{center}
			\includegraphics[width=\columnwidth]{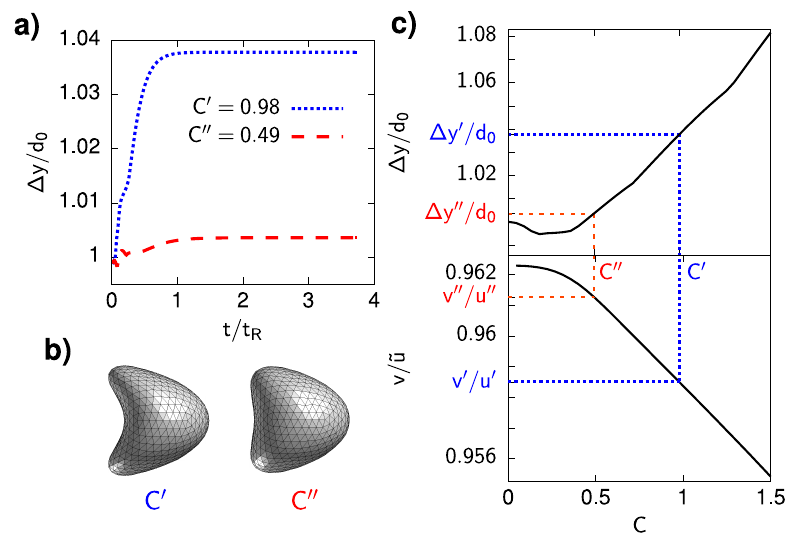}
		\end{center}
	\vspace{-0.4cm}
		\caption{Shape-dependent entrainment of capsules in stationary Poiseuille flow.
		 a): Time evolution of the capsule's lateral extension $\Delta y$ after the onset of the flow at $t=0$ for the two capillary numbers $C^{\prime}=0.98$ and $C^{\prime\prime}=0.49$. 
		 b): Simulation snapshots of the capsule after the stationary shape is reached.
		 c): Plateau value of the lateral size [see a)] (top) and relative velocity $v/\tilde u$ (bottom) as function of the capillary number. Increasing $\Delta y$ results in decreasing $v/\tilde{u}$.
 }
		\label{fig_calibration_capsule}
	\end{figure}

Whereas the vertical extend of the MM and the RBC decrease with $\tilde u$, the lateral size of the capsule and the ring polymer increases with $C$.
In Fig.~\ref{fig_calibration_capsule}a) we show the evolution of the deformation of a capsule after a sudden onset of a stationary parabolic flow for the two values $C^{\prime}=0.98$ and $C^{\prime\prime}=0.49$. After a time of the order of $t_R$, the capsule's deformation reaches a stationary, bullet-like shape with a vertical  extension larger than in its undeformed state, i.e.  $\Delta y^{\prime}>\Delta y^{\prime\prime}>d_0$. The corresponding snapshots are shown in Fig.~\ref{fig_calibration_capsule}b). Similar deformations are found for stiffer cells \cite{GuckJ:2015.1,GuckJ:2015.2}. This increase in lateral size is in contrast to the RBC and in agreement with previous simulation results on capsules  \cite{Villone:2016.1}.

Note that any finite-sized particle in a parabolic flow profile lags behind $\tilde u$, i.\ e.\ $v / \tilde{u}<1$ always holds.
Using the same reasoning as above, an increasing vertical extent $\Delta y$  for capsules leads to a decreasing ratio
 $v/\tilde u$, as shown 
in Fig.~\ref{fig_calibration_capsule}c).
In the asymmetrically oscillating flow as given in Eq.\,(\ref{eq_background_flow}), this results in $ v_1/\tilde u_1< |v_2/\tilde u_2|$ and a negative propulsion step for the capsule and the ring polymer.

The determination of the evolution of a soft particle's deformation  such as for the 
capsule in Fig.\,\ref{fig_calibration_capsule}a) gives the lower bounds of $T_{1,2}$
for a pulsating flow
as well as  the sign of $\Delta x$. $T_{1,2}$  should always be chosen significantly larger than $t_R$  of a soft particle in order to obtain a reasonable difference between the traveled distances per flow section, $v_1 T_1$ and $|v_2| T_2$. 
Since $\Delta x \propto T$ according to Eq.\ (\ref{eq_analytic_actuation_velocity_def}), the total net progress after $n$ oscillation cycles will be the same if $T$ is increased and $n$ decreased proportionally, provided that $t_R \ll T_{1,2}$ applies.

\begin{figure}[htb]
	\begin{center}
		\includegraphics[width=\columnwidth]{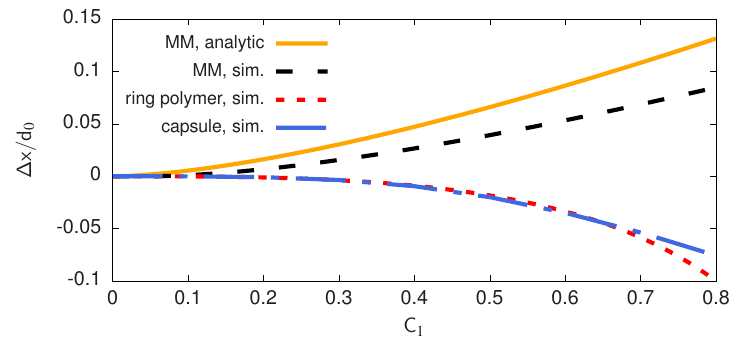}
	\end{center}
	\vspace{-0.4cm}
	\caption{Propulsion step $\Delta x$ as a function of the capillary number of the forward flow section, $C_1$, for different particles in unbounded flows: Capsule (blue long-dashed), ring polymer (red short-dashed) and minimal model (analytical: orange solid, simulation: black dashed). $\Delta x$ is given in units of the respective initial particle diameters $d_0$.  
	}
	\label{fig_flow_amplitude}
\end{figure}

The different sign and the dependence of $\Delta x$  on $C_1$ is shown in Fig.\,\ref{fig_flow_amplitude} for  three different  particles.  $C_1$ is changed by the flow amplitude $\tilde u$ in  $T_{1,2}$ with $C_2=C_1/A$ and $A=2$.
For all particles, the magnitude $|\Delta x|$ increases with growing $C_1$.  The analytical result  according to Eq.~(\ref{eq_analytic_actuation_velocity_final}) approximates the numerical results for the MM   in Fig.\,\ref{fig_flow_amplitude}  quite well, especially for small $C_1$.

\begin{figure}[htb]
	\begin{center}
		\includegraphics[width=\columnwidth]{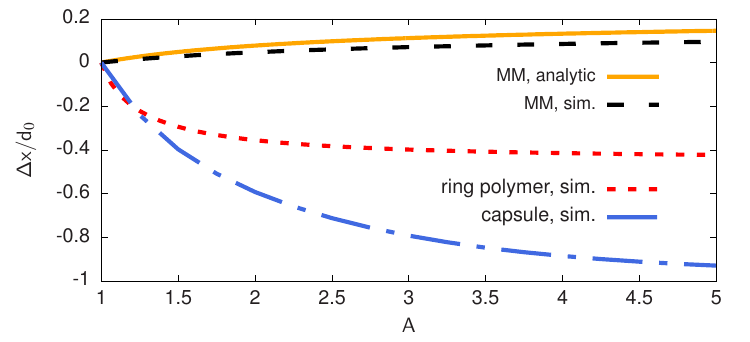}
	\end{center}
	\vspace{-0.4cm}
	\caption{Propulsion step $\Delta x$ in units of the initial particle diameter $d_0$ as a function of the asymmetry of the oscillating flow, $A$, for
		the same particles as in Fig.\,\ref{fig_flow_amplitude}.
	}
	\label{fig_flow_asymmetry}
\end{figure}

$|\Delta x|$ increases monotonically with $A$ for all types of soft particles, as shown in
Fig.\,\ref{fig_flow_asymmetry}. 
Starting from $A=1$, we increase the flow asymmetry by lengthening $T_2$ and decreasing $u_2$ accordingly, while keeping $T_1$ and $u_1$ fixed. The analytical  approximation [see Eq.\,(\ref{eq_analytic_actuation_velocity_final})] and numerical calculations for the MM also agree well here.

\begin{figure}[htb]
		\begin{center}
			\includegraphics[width=\columnwidth]{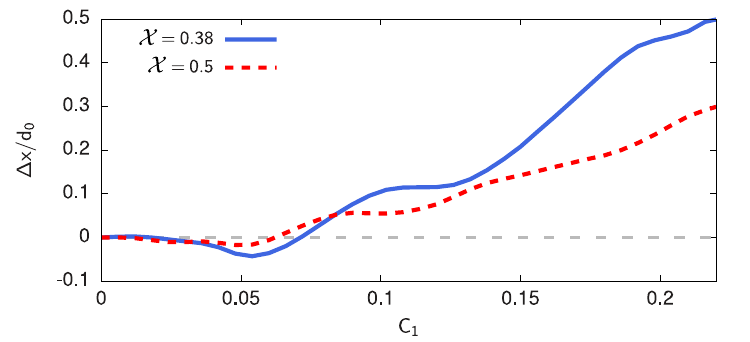}
		\end{center}
	\vspace{-0.4cm}
		\caption{Propulsion step $\Delta x$  of a red blood cell in units of its initial diameter $d_0$ as function of the capillary number $C_1$ in the forward section of the asymmetrically oscillating flow. The results are obtained by Lattice-Boltzmann simulations (bounded flows) with a flow asymmetry $A=4$ for two confinements $\chi = 0.38$ (solid) and $\chi = 0.5$ (dashed).
 }
		\label{fig_rbc_lbm}
	\end{figure}

So far our results for simulations in unbounded flows show that the net propulsion of soft particles in asymmetrically oscillating flows originates from a leading bulk contribution as the particle-wall HI has been neglected.
This is a good approximation for the case of a small confinement parameter $\chi=d_0/(2w)$.
In order to address the influence of the walls, we turn towards LBM simulations in the following. In Fig.\,\ref{fig_rbc_lbm} the propulsion step per flow period is shown as function of the capillary number for a RBC in bounded Poiseuille flow with $A=4$. The results for two different, experimentally common values of $\chi=0.5,0.38$ are displayed. For the determination of $\Delta x$, we take the average value covering the range of three flow periods.   
For small $C_1$, $\Delta x$ becomes negative at first, similar to capsules. However, the propulsion step changes sign at intermediate values of $C_1$ and then continues to grow monotonically with $C_1$.
Nevertheless, the magnitude $|\Delta x|$ is larger for $\chi=0.38$ than for $\chi=0.5$ throughout the majority of values of the capillary number in Fig.\,\ref{fig_rbc_lbm}. 
Therefore we conclude that the propulsive effect is reduced, but only quantitatively changed by the particle-wall HI. For a multi-particle simulation we refer to Ref.\ \cite{supplement}. Our results are confirmed by recent experiments on RBCs in pulsating flows \cite{krauss}.


In this work, we identified and described a new propulsion mechanism for soft particles in microchannels. 
In contrast to so-called microfluidic deterministic ratchets \cite{Austin:2009.1,McFaul:2012.1,ParkE:2016.1} it does not rely on asymmetric posts in microchannels for directed particle motion in pulsating flows. Rather, time-reversal symmetry is broken by using different
 forward and backward velocities of an on average non-progressing liquid.  
Since soft particles are  deformed and entrained differently in the two flow sections, they are selectively propelled according to their deformability, whereas rigid particles do not move on average.

The propulsion direction of a soft particle in  pulsating microflows depends on its deformation type. For example, the lateral extension of an originally biconcave RBC becomes smaller in a Poiseuille flow. In contrast, the vertical extension increases for capsules or Hela cells in Ref.\,\cite{GuckJ:2015.1}. The advance step has opposite sign  but particles of different elasticity are separated in both cases.

The presented generic propulsion phenomenon is independent 
on specific elastic  properties of soft particles.
Another great advantage of the method is that even cells with small  elasticity differences  can be separated by increasing the number of forward/reverse flow cycles. 
Furthermore, the oscillating fluid motion allows an efficient particle separation even in short channels compared to other methods.
Since the suggested particle propulsion mechanism operates in the range of small Reynolds numbers (and thus small flow amplitudes), even larger bio-particles such as  circulating tumor cells (CTCs) 
can be sorted at physiological shear stresses without the cells being damaged or showing a flow-induced response.

Our robust and label-free method fulfills a great need for separating cells with different mechanical properties, e.g.\ malignant cells and their healthy counterparts.
Assuming RBCs of size $d_0 = 8 \mu \text{m}$ with a shear modulus $\kappa_\text{S}$ of $7 \mu \text{Nm}^{-1}$ for healthy cells and $10 \mu \text{Nm}^{-1}$ for RBCs with the sickle cell mutation \cite{SaadS:2003.1, suresh_2006}, we can approximate their different net progress. After $250 \text{s}$ of oscillation with a flow amplitude of $u_1 = 4.8 \text{mms}^{-1}$ and a flow asymmetry of $4$ in a microchannel with height $2w = 21 \mu \text{m}$, healthy RBCs will have moved $3.6 \text{mm}$, malignant cells  only by $1.3 \text{mm}$ on average.


We gratefully acknowledge discussions with P.-Y.~Gires, S.W.~Krauss, and M.~Weiss.
For support W.S. thanks the DAAD, W.S. and W.Z. the Elite Study Program Biological Physics, and all authors the French-German university (Grant No. CFDA-Q1-14, “Living fluids”).


\begin{thebibliography}{68}%
	\makeatletter
	\providecommand \@ifxundefined [1]{%
		\@ifx{#1\undefined}
	}%
	\providecommand \@ifnum [1]{%
		\ifnum #1\expandafter \@firstoftwo
		\else \expandafter \@secondoftwo
		\fi
	}%
	\providecommand \@ifx [1]{%
		\ifx #1\expandafter \@firstoftwo
		\else \expandafter \@secondoftwo
		\fi
	}%
	\providecommand \natexlab [1]{#1}%
	\providecommand \enquote  [1]{``#1''}%
	\providecommand \bibnamefont  [1]{#1}%
	\providecommand \bibfnamefont [1]{#1}%
	\providecommand \citenamefont [1]{#1}%
	\providecommand \href@noop [0]{\@secondoftwo}%
	\providecommand \href [0]{\begingroup \@sanitize@url \@href}%
	\providecommand \@href[1]{\@@startlink{#1}\@@href}%
	\providecommand \@@href[1]{\endgroup#1\@@endlink}%
	\providecommand \@sanitize@url [0]{\catcode `\\12\catcode `\$12\catcode
		`\&12\catcode `\#12\catcode `\^12\catcode `\_12\catcode `\%12\relax}%
	\providecommand \@@startlink[1]{}%
	\providecommand \@@endlink[0]{}%
	\providecommand \url  [0]{\begingroup\@sanitize@url \@url }%
	\providecommand \@url [1]{\endgroup\@href {#1}{\urlprefix }}%
	\providecommand \urlprefix  [0]{URL }%
	\providecommand \Eprint [0]{\href }%
	\providecommand \doibase [0]{http://dx.doi.org/}%
	\providecommand \selectlanguage [0]{\@gobble}%
	\providecommand \bibinfo  [0]{\@secondoftwo}%
	\providecommand \bibfield  [0]{\@secondoftwo}%
	\providecommand \translation [1]{[#1]}%
	\providecommand \BibitemOpen [0]{}%
	\providecommand \bibitemStop [0]{}%
	\providecommand \bibitemNoStop [0]{.\EOS\space}%
	\providecommand \EOS [0]{\spacefactor3000\relax}%
	\providecommand \BibitemShut  [1]{\csname bibitem#1\endcsname}%
	\let\auto@bib@innerbib\@empty
	\bibitem [{\citenamefont {Lee}\ \emph {et~al.}(2017)\citenamefont {Lee},
		\citenamefont {Tseng},\ and\ \citenamefont {Carlo}}]{DiCarlo:2017}%
	\BibitemOpen
	\bibinfo {editor} {\bibfnamefont {W.}~\bibnamefont {Lee}}, \bibinfo {editor}
	{\bibfnamefont {P.}~\bibnamefont {Tseng}}, \ and\ \bibinfo {editor}
	{\bibfnamefont {D.~Di}\ \bibnamefont {Carlo}},\ eds.,\ \href@noop {} {\emph
		{\bibinfo {title} {Microtechnology for Cell Manipulation and Sorting}}}\
	(\bibinfo  {publisher} {Springer},\ \bibinfo {address} {Cham, Switzerland},\
	\bibinfo {year} {2017})\BibitemShut {NoStop}%
	\bibitem [{\citenamefont {Nguyen}\ \emph {et~al.}(2019)\citenamefont {Nguyen},
		\citenamefont {Wereley},\ and\ \citenamefont {Shaegh}}]{Nguyen:2019}%
	\BibitemOpen
	\bibfield  {author} {\bibinfo {author} {\bibfnamefont {N.-T.}\ \bibnamefont
			{Nguyen}}, \bibinfo {author} {\bibfnamefont {S.~T.}\ \bibnamefont {Wereley}},
		\ and\ \bibinfo {author} {\bibfnamefont {S.~A.~M.}\ \bibnamefont {Shaegh}},\
	}\href@noop {} {\emph {\bibinfo {title} {Fundamentals and Applications of
				Microfluidics}}}\ (\bibinfo  {publisher} {Artech House},\ \bibinfo {address}
	{Boston},\ \bibinfo {year} {2019})\BibitemShut {NoStop}%
	\bibitem [{\citenamefont {Bhagat}\ \emph {et~al.}(2010)\citenamefont {Bhagat},
		\citenamefont {Bow}, \citenamefont {Hou}, \citenamefont {Tan}, \citenamefont
		{Han},\ and\ \citenamefont {Lim}}]{LimCT:2010.2}%
	\BibitemOpen
	\bibfield  {author} {\bibinfo {author} {\bibfnamefont {A.~A.~S.}\
			\bibnamefont {Bhagat}}, \bibinfo {author} {\bibfnamefont {H.}~\bibnamefont
			{Bow}}, \bibinfo {author} {\bibfnamefont {H.~W.}\ \bibnamefont {Hou}},
		\bibinfo {author} {\bibfnamefont {S.~J.}\ \bibnamefont {Tan}}, \bibinfo
		{author} {\bibfnamefont {J.}~\bibnamefont {Han}}, \ and\ \bibinfo {author}
		{\bibfnamefont {C.~T.}\ \bibnamefont {Lim}},\ }\bibfield  {title} {\enquote
		{\bibinfo {title} {Microfluidics for cell separation},}\ }\href@noop {}
	{\bibfield  {journal} {\bibinfo  {journal} {Med. Biol. Eng. Comput.}\
		}\textbf {\bibinfo {volume} {48}},\ \bibinfo {pages} {999} (\bibinfo {year}
		{2010})}\BibitemShut {NoStop}%
	\bibitem [{\citenamefont {Karimi}\ \emph {et~al.}(2013)\citenamefont {Karimi},
		\citenamefont {Yazdi},\ and\ \citenamefont {Ardekani}}]{Karimi:2013.1}%
	\BibitemOpen
	\bibfield  {author} {\bibinfo {author} {\bibfnamefont {A.}~\bibnamefont
			{Karimi}}, \bibinfo {author} {\bibfnamefont {S.}~\bibnamefont {Yazdi}}, \
		and\ \bibinfo {author} {\bibfnamefont {A.~M.}\ \bibnamefont {Ardekani}},\
	}\bibfield  {title} {\enquote {\bibinfo {title} {Hydrodynamic mechanisms of
				cell and particle trapping in microfluidics},}\ }\href@noop {} {\bibfield
		{journal} {\bibinfo  {journal} {Biomicrofluidics}\ }\textbf {\bibinfo
			{volume} {7}},\ \bibinfo {pages} {021501} (\bibinfo {year}
		{2013})}\BibitemShut {NoStop}%
	\bibitem [{\citenamefont {Sajeesh}\ and\ \citenamefont
		{Sen}(2014)}]{Sajeesh:2014.1}%
	\BibitemOpen
	\bibfield  {author} {\bibinfo {author} {\bibfnamefont {P.}~\bibnamefont
			{Sajeesh}}\ and\ \bibinfo {author} {\bibfnamefont {A.~K.}\ \bibnamefont
			{Sen}},\ }\bibfield  {title} {\enquote {\bibinfo {title} {{Particle
					separation and sorting in microfluidic devices: a review}},}\ }\href@noop {}
	{\bibfield  {journal} {\bibinfo  {journal} {Microfluid Nanofluidics}\
		}\textbf {\bibinfo {volume} {17}},\ \bibinfo {pages} {1} (\bibinfo {year}
		{2014})}\BibitemShut {NoStop}%
	\bibitem [{\citenamefont {{Shields IV}}\ \emph {et~al.}(2015)\citenamefont
		{{Shields IV}}, \citenamefont {Reyes},\ and\ \citenamefont
		{L\'opez}}]{LopezG:2015.1}%
	\BibitemOpen
	\bibfield  {author} {\bibinfo {author} {\bibfnamefont {C.~W.}\ \bibnamefont
			{{Shields IV}}}, \bibinfo {author} {\bibfnamefont {C.~D.}\ \bibnamefont
			{Reyes}}, \ and\ \bibinfo {author} {\bibfnamefont {G.~P.}\ \bibnamefont
			{L\'opez}},\ }\bibfield  {title} {\enquote {\bibinfo {title} {Microfluidic
				cell sorting: a review of the advances in the separation of cells from
				debulking to rare cell isolation},}\ }\href {\doibase 10.1039/C4LC01246A}
	{\bibfield  {journal} {\bibinfo  {journal} {Lab Chip}\ }\textbf {\bibinfo
			{volume} {15}},\ \bibinfo {pages} {1230} (\bibinfo {year}
		{2015})}\BibitemShut {NoStop}%
	\bibitem [{\citenamefont {Dahl}\ \emph {et~al.}(2015)\citenamefont {Dahl},
		\citenamefont {Lin}, \citenamefont {Muller},\ and\ \citenamefont
		{Kumar}}]{Kumar_S:2015.1}%
	\BibitemOpen
	\bibfield  {author} {\bibinfo {author} {\bibfnamefont {J.~B.}\ \bibnamefont
			{Dahl}}, \bibinfo {author} {\bibfnamefont {J.-M.~G.}\ \bibnamefont {Lin}},
		\bibinfo {author} {\bibfnamefont {S.~J.}\ \bibnamefont {Muller}}, \ and\
		\bibinfo {author} {\bibfnamefont {S.}~\bibnamefont {Kumar}},\ }\bibfield
	{title} {\enquote {\bibinfo {title} {Microfluidic strategies for
				understanding the mechanics of cells and cell-mimetic systems},}\ }\href@noop
	{} {\bibfield  {journal} {\bibinfo  {journal} {Annu. Rev. Chem. Biomol.
				Eng.}\ }\textbf {\bibinfo {volume} {6}},\ \bibinfo {pages} {293} (\bibinfo
		{year} {2015})}\BibitemShut {NoStop}%
	\bibitem [{\citenamefont {Lee}\ \emph {et~al.}(2016)\citenamefont {Lee},
		\citenamefont {Kim}, \citenamefont {Ahn}, \citenamefont {Lee},\ and\
		\citenamefont {Park}}]{ParkJY:2016.1}%
	\BibitemOpen
	\bibfield  {author} {\bibinfo {author} {\bibfnamefont {G.-H.}\ \bibnamefont
			{Lee}}, \bibinfo {author} {\bibfnamefont {S.-H.}\ \bibnamefont {Kim}},
		\bibinfo {author} {\bibfnamefont {K.}~\bibnamefont {Ahn}}, \bibinfo {author}
		{\bibfnamefont {S.-H.}\ \bibnamefont {Lee}}, \ and\ \bibinfo {author}
		{\bibfnamefont {J.~Y.}\ \bibnamefont {Park}},\ }\bibfield  {title} {\enquote
		{\bibinfo {title} {{Separation and sorting of cells in microsystems using
					physical principles}},}\ }\href@noop {} {\bibfield  {journal} {\bibinfo
			{journal} {J. Micromech. Microeng.}\ }\textbf {\bibinfo {volume} {26}},\
		\bibinfo {pages} {013003} (\bibinfo {year} {2016})}\BibitemShut {NoStop}%
	\bibitem [{\citenamefont {Stoecklein}\ and\ \citenamefont
		{Carlo}(2019)}]{DiCarlo:2019.1}%
	\BibitemOpen
	\bibfield  {author} {\bibinfo {author} {\bibfnamefont {D.}~\bibnamefont
			{Stoecklein}}\ and\ \bibinfo {author} {\bibfnamefont {D.~Di}\ \bibnamefont
			{Carlo}},\ }\bibfield  {title} {\enquote {\bibinfo {title} {{Nonlinear
					Microfluidics}},}\ }\href@noop {} {\bibfield  {journal} {\bibinfo  {journal}
			{Anal. Chem.}\ }\textbf {\bibinfo {volume} {91}},\ \bibinfo {pages} {296}
		(\bibinfo {year} {2019})}\BibitemShut {NoStop}%
	\bibitem [{\citenamefont {Nasiri}\ \emph {et~al.}(2020)\citenamefont {Nasiri},
		\citenamefont {Shamloo}, \citenamefont {Ahadian}, \citenamefont {Amirifar},
		\citenamefont {Akbari}, \citenamefont {Goudie}, \citenamefont {Ashammakhi},
		\citenamefont {Dokmeci}, \citenamefont {DiCarlo},\ and\ \citenamefont
		{Khademhosseini}}]{Nasiri:2020.1}%
	\BibitemOpen
	\bibfield  {author} {\bibinfo {author} {\bibfnamefont {R.}~\bibnamefont
			{Nasiri}}, \bibinfo {author} {\bibfnamefont {A.}~\bibnamefont {Shamloo}},
		\bibinfo {author} {\bibfnamefont {S.}~\bibnamefont {Ahadian}}, \bibinfo
		{author} {\bibfnamefont {L.}~\bibnamefont {Amirifar}}, \bibinfo {author}
		{\bibfnamefont {J.}~\bibnamefont {Akbari}}, \bibinfo {author} {\bibfnamefont
			{M.~J.}\ \bibnamefont {Goudie}}, \bibinfo {author} {\bibfnamefont
			{K.~Lee~N.}\ \bibnamefont {Ashammakhi}}, \bibinfo {author} {\bibfnamefont
			{M.~R.}\ \bibnamefont {Dokmeci}}, \bibinfo {author} {\bibfnamefont
			{D.}~\bibnamefont {DiCarlo}}, \ and\ \bibinfo {author} {\bibfnamefont
			{A.}~\bibnamefont {Khademhosseini}},\ }\bibfield  {title} {\enquote {\bibinfo
			{title} {Microfluidic-based approaches and targeted cell/particle separation
				based on physical properties: Fundamentals and applications},}\ }\href@noop
	{} {\bibfield  {journal} {\bibinfo  {journal} {Small}\ }\textbf {\bibinfo
			{volume} {16}},\ \bibinfo {pages} {2000171} (\bibinfo {year}
		{2020})}\BibitemShut {NoStop}%
	\bibitem [{\citenamefont {Zhang}\ \emph {et~al.}(2020)\citenamefont {Zhang},
		\citenamefont {Wang}, \citenamefont {Onck},\ and\ \citenamefont {den
			Toonder}}]{Toonder:2020.1}%
	\BibitemOpen
	\bibfield  {author} {\bibinfo {author} {\bibfnamefont {S.}~\bibnamefont
			{Zhang}}, \bibinfo {author} {\bibfnamefont {Y.}~\bibnamefont {Wang}},
		\bibinfo {author} {\bibfnamefont {P.}~\bibnamefont {Onck}}, \ and\ \bibinfo
		{author} {\bibfnamefont {J.}~\bibnamefont {den Toonder}},\ }\bibfield
	{title} {\enquote {\bibinfo {title} {A concise review of microfluidic
				particle manipulation methods},}\ }\href@noop {} {\bibfield  {journal}
		{\bibinfo  {journal} {Microfluid Nanofluidics}\ }\textbf {\bibinfo {volume}
			{24}},\ \bibinfo {pages} {24} (\bibinfo {year} {2020})}\BibitemShut {NoStop}%
	\bibitem [{\citenamefont {Lin}\ \emph {et~al.}(2020)\citenamefont {Lin},
		\citenamefont {Luo}, \citenamefont {Du}, \citenamefont {Kong}, \citenamefont
		{Liu},\ and\ \citenamefont {Liu}}]{LiuZ:2020.1}%
	\BibitemOpen
	\bibfield  {author} {\bibinfo {author} {\bibfnamefont {Z.}~\bibnamefont
			{Lin}}, \bibinfo {author} {\bibfnamefont {G.}~\bibnamefont {Luo}}, \bibinfo
		{author} {\bibfnamefont {W.}~\bibnamefont {Du}}, \bibinfo {author}
		{\bibfnamefont {T.}~\bibnamefont {Kong}}, \bibinfo {author} {\bibfnamefont
			{C.}~\bibnamefont {Liu}}, \ and\ \bibinfo {author} {\bibfnamefont
			{Z.}~\bibnamefont {Liu}},\ }\bibfield  {title} {\enquote {\bibinfo {title}
			{{Recent Advances in Microfluidic Platforms Applied in Cancer Metastasis:
					Circulating Tumor Cells' (CTCs) Isolation and Tumor-On-A-Chip}},}\
	}\href@noop {} {\bibfield  {journal} {\bibinfo  {journal} {Small}\ }\textbf
		{\bibinfo {volume} {16}},\ \bibinfo {pages} {1903899} (\bibinfo {year}
		{2020})}\BibitemShut {NoStop}%
	\bibitem [{\citenamefont {Suresh}(2007)}]{SURESH2007413}%
	\BibitemOpen
	\bibfield  {author} {\bibinfo {author} {\bibfnamefont {S.}~\bibnamefont
			{Suresh}},\ }\bibfield  {title} {\enquote {\bibinfo {title} {{Biomechanics
					and biophysics of cancer cells}},}\ }\href@noop {} {\bibfield  {journal}
		{\bibinfo  {journal} {Acta Biomater.}\ }\textbf {\bibinfo {volume} {3}},\
		\bibinfo {pages} {413} (\bibinfo {year} {2007})}\BibitemShut {NoStop}%
	\bibitem [{\citenamefont {{J. Guck \textit{et al.}}}(2005)}]{GUCK20053689}%
	\BibitemOpen
	\bibfield  {author} {\bibinfo {author} {\bibnamefont {{J. Guck \textit{et
						al.}}}},\ }\bibfield  {title} {\enquote {\bibinfo {title} {{Optical
					Deformability as an Inherent Cell Marker for Testing Malignant Transformation
					and Metastatic Competence}},}\ }\href@noop {} {\bibfield  {journal} {\bibinfo
			{journal} {Biophys. J.}\ }\textbf {\bibinfo {volume} {88}},\ \bibinfo
		{pages} {3689} (\bibinfo {year} {2005})}\BibitemShut {NoStop}%
	\bibitem [{\citenamefont {Brand{\~{a}}o}\ \emph {et~al.}(2003)\citenamefont
		{Brand{\~{a}}o}, \citenamefont {Fontes}, \citenamefont {Barjas-Castro},
		\citenamefont {Barbosa}, \citenamefont {Costa}, \citenamefont {Cesar},\ and\
		\citenamefont {Saad}}]{SaadS:2003.1}%
	\BibitemOpen
	\bibfield  {author} {\bibinfo {author} {\bibfnamefont {M.~M.}\ \bibnamefont
			{Brand{\~{a}}o}}, \bibinfo {author} {\bibfnamefont {A.}~\bibnamefont
			{Fontes}}, \bibinfo {author} {\bibfnamefont {M.~L.}\ \bibnamefont
			{Barjas-Castro}}, \bibinfo {author} {\bibfnamefont {L.~C.}\ \bibnamefont
			{Barbosa}}, \bibinfo {author} {\bibfnamefont {F.~F.}\ \bibnamefont {Costa}},
		\bibinfo {author} {\bibfnamefont {C.~L.}\ \bibnamefont {Cesar}}, \ and\
		\bibinfo {author} {\bibfnamefont {S.~T.~O.}\ \bibnamefont {Saad}},\
	}\bibfield  {title} {\enquote {\bibinfo {title} {Optical tweezers for
				measuring red blood cell elasticity: application to the study of drug
				response in sickle cell disease},}\ }\href@noop {} {\bibfield  {journal}
		{\bibinfo  {journal} {Eur. J. Haematol.}\ }\textbf {\bibinfo {volume} {70}},\
		\bibinfo {pages} {207} (\bibinfo {year} {2003})}\BibitemShut {NoStop}%
	\bibitem [{\citenamefont {Drost}\ and\ \citenamefont
		{MacNee}(2002)}]{MacNee:2002.1}%
	\BibitemOpen
	\bibfield  {author} {\bibinfo {author} {\bibfnamefont {E.~M.}\ \bibnamefont
			{Drost}}\ and\ \bibinfo {author} {\bibfnamefont {W.}~\bibnamefont {MacNee}},\
	}\bibfield  {title} {\enquote {\bibinfo {title} {{Potential role of IL-8,
					platelet-activating factor and TNF-$\alpha$ in the sequestration of
					neutrphils in the lung: effects of neutrophil deformability, adhesion
					receptor expression, and chemotaxis}},}\ }\href@noop {} {\bibfield  {journal}
		{\bibinfo  {journal} {Eur. J. Immunol.}\ }\textbf {\bibinfo {volume} {32}},\
		\bibinfo {pages} {393} (\bibinfo {year} {2002})}\BibitemShut {NoStop}%
	\bibitem [{\citenamefont {Dondorp}\ \emph {et~al.}(2000)\citenamefont
		{Dondorp}, \citenamefont {Kager}, \citenamefont {Vreeken},\ and\
		\citenamefont {White}}]{Dondrop:2000.1}%
	\BibitemOpen
	\bibfield  {author} {\bibinfo {author} {\bibfnamefont {A.~M.}\ \bibnamefont
			{Dondorp}}, \bibinfo {author} {\bibfnamefont {P.~A.}\ \bibnamefont {Kager}},
		\bibinfo {author} {\bibfnamefont {J.}~\bibnamefont {Vreeken}}, \ and\
		\bibinfo {author} {\bibfnamefont {N.~J.}\ \bibnamefont {White}},\ }\bibfield
	{title} {\enquote {\bibinfo {title} {Abnormal blood flow and red blood cell
				deformability in severe malaria},}\ }\href@noop {} {\bibfield  {journal}
		{\bibinfo  {journal} {Parasitol. Today}\ }\textbf {\bibinfo {volume} {16}},\
		\bibinfo {pages} {228} (\bibinfo {year} {2000})}\BibitemShut {NoStop}%
	\bibitem [{\citenamefont {McMillan}\ \emph {et~al.}(1978)\citenamefont
		{McMillan}, \citenamefont {Utterback},\ and\ \citenamefont
		{Puma}}]{McMillan:1978.1}%
	\BibitemOpen
	\bibfield  {author} {\bibinfo {author} {\bibfnamefont {D.~E.}\ \bibnamefont
			{McMillan}}, \bibinfo {author} {\bibfnamefont {N.~G.}\ \bibnamefont
			{Utterback}}, \ and\ \bibinfo {author} {\bibfnamefont {J.~La}\ \bibnamefont
			{Puma}},\ }\bibfield  {title} {\enquote {\bibinfo {title} {Reduced
				erythrocyte deformability in diabetes},}\ }\href@noop {} {\bibfield
		{journal} {\bibinfo  {journal} {Diabetes}\ }\textbf {\bibinfo {volume}
			{27}},\ \bibinfo {pages} {895} (\bibinfo {year} {1978})}\BibitemShut
	{NoStop}%
	\bibitem [{\citenamefont {Debnath}\ and\ \citenamefont
		{Sadrzadeh}(2018)}]{Debnath:2018.1}%
	\BibitemOpen
	\bibfield  {author} {\bibinfo {author} {\bibfnamefont {N.}~\bibnamefont
			{Debnath}}\ and\ \bibinfo {author} {\bibfnamefont {M.}~\bibnamefont
			{Sadrzadeh}},\ }\bibfield  {title} {\enquote {\bibinfo {title} {{Microfluidic
					Mimic for Colloid Membrane Filtration: A Review}},}\ }\href@noop {}
	{\bibfield  {journal} {\bibinfo  {journal} {J. Indian Inst. Sci.}\ }\textbf
		{\bibinfo {volume} {98}},\ \bibinfo {pages} {137} (\bibinfo {year}
		{2018})}\BibitemShut {NoStop}%
	\bibitem [{\citenamefont {Zhang}\ \emph {et~al.}(2016)\citenamefont {Zhang},
		\citenamefont {Yan}, \citenamefont {Yuan}, \citenamefont {Alici},
		\citenamefont {Nguyen}, \citenamefont {Warkiani},\ and\ \citenamefont
		{Li}}]{LiWeihua:2016.1}%
	\BibitemOpen
	\bibfield  {author} {\bibinfo {author} {\bibfnamefont {J.}~\bibnamefont
			{Zhang}}, \bibinfo {author} {\bibfnamefont {S.}~\bibnamefont {Yan}}, \bibinfo
		{author} {\bibfnamefont {D.}~\bibnamefont {Yuan}}, \bibinfo {author}
		{\bibfnamefont {G.}~\bibnamefont {Alici}}, \bibinfo {author} {\bibfnamefont
			{N.-T.}\ \bibnamefont {Nguyen}}, \bibinfo {author} {\bibfnamefont {M.~E.}\
			\bibnamefont {Warkiani}}, \ and\ \bibinfo {author} {\bibfnamefont
			{W.}~\bibnamefont {Li}},\ }\bibfield  {title} {\enquote {\bibinfo {title}
			{Fundamentals and applications of inertial microfluidics: a review},}\
	}\href@noop {} {\bibfield  {journal} {\bibinfo  {journal} {Lab Chip}\
		}\textbf {\bibinfo {volume} {16}},\ \bibinfo {pages} {10} (\bibinfo {year}
		{2016})}\BibitemShut {NoStop}%
	\bibitem [{\citenamefont {Huang}\ \emph {et~al.}(2004)\citenamefont {Huang},
		\citenamefont {Cox}, \citenamefont {Austin},\ and\ \citenamefont
		{Sturm}}]{Austin:2004.1}%
	\BibitemOpen
	\bibfield  {author} {\bibinfo {author} {\bibfnamefont {L.~R.}\ \bibnamefont
			{Huang}}, \bibinfo {author} {\bibfnamefont {E.~C.}\ \bibnamefont {Cox}},
		\bibinfo {author} {\bibfnamefont {R.~H.}\ \bibnamefont {Austin}}, \ and\
		\bibinfo {author} {\bibfnamefont {J.~C.}\ \bibnamefont {Sturm}},\ }\bibfield
	{title} {\enquote {\bibinfo {title} {Continuous particle separation through
				deterministic lateral displacement},}\ }\href@noop {} {\bibfield  {journal}
		{\bibinfo  {journal} {Science}\ }\textbf {\bibinfo {volume} {304}},\ \bibinfo
		{pages} {987} (\bibinfo {year} {2004})}\BibitemShut {NoStop}%
	\bibitem [{\citenamefont {Inglis}\ \emph {et~al.}(2006)\citenamefont {Inglis},
		\citenamefont {Davis}, \citenamefont {Austin},\ and\ \citenamefont
		{Sturm}}]{Austin:2006.2}%
	\BibitemOpen
	\bibfield  {author} {\bibinfo {author} {\bibfnamefont {D.~W.}\ \bibnamefont
			{Inglis}}, \bibinfo {author} {\bibfnamefont {J.~A.}\ \bibnamefont {Davis}},
		\bibinfo {author} {\bibfnamefont {R.~H.}\ \bibnamefont {Austin}}, \ and\
		\bibinfo {author} {\bibfnamefont {J.~C.}\ \bibnamefont {Sturm}},\ }\bibfield
	{title} {\enquote {\bibinfo {title} {Critical particle size for fractionation
				by deterministic lateral displacement},}\ }\href@noop {} {\bibfield
		{journal} {\bibinfo  {journal} {Lab Chip}\ }\textbf {\bibinfo {volume} {6}},\
		\bibinfo {pages} {655} (\bibinfo {year} {2006})}\BibitemShut {NoStop}%
	\bibitem [{\citenamefont {McGrath}\ \emph {et~al.}(2014)\citenamefont
		{McGrath}, \citenamefont {Jimenez},\ and\ \citenamefont
		{Bridle}}]{Bridle:2014.1}%
	\BibitemOpen
	\bibfield  {author} {\bibinfo {author} {\bibfnamefont {J.}~\bibnamefont
			{McGrath}}, \bibinfo {author} {\bibfnamefont {M.}~\bibnamefont {Jimenez}}, \
		and\ \bibinfo {author} {\bibfnamefont {H.}~\bibnamefont {Bridle}},\
	}\bibfield  {title} {\enquote {\bibinfo {title} {{Deterministic lateral
					displacement for particle separation: a review}},}\ }\href@noop {} {\bibfield
		{journal} {\bibinfo  {journal} {Lab Chip}\ }\textbf {\bibinfo {volume}
			{14}},\ \bibinfo {pages} {4139} (\bibinfo {year} {2014})}\BibitemShut
	{NoStop}%
	\bibitem [{\citenamefont {{A. Hochstetter \textit{et
					al.}}}(2020)}]{AustinGomp:2020.1}%
	\BibitemOpen
	\bibfield  {author} {\bibinfo {author} {\bibnamefont {{A. Hochstetter
					\textit{et al.}}}},\ }\bibfield  {title} {\enquote {\bibinfo {title}
			{{Deterministic Lateral Displacement: Challenges and Perspectives}},}\
	}\href@noop {} {\bibfield  {journal} {\bibinfo  {journal} {ACS Nano}\
		}\textbf {\bibinfo {volume} {14}},\ \bibinfo {pages} {10784} (\bibinfo {year}
		{2020})}\BibitemShut {NoStop}%
	\bibitem [{\citenamefont {{Z. Zhang and W. Chien and E. Henry and D.~A. Fedosov
				and G. Gompper}}(2019)}]{Gompper:2019.1}%
	\BibitemOpen
	\bibfield  {author} {\bibinfo {author} {\bibnamefont {{Z. Zhang and W. Chien
					and E. Henry and D.~A. Fedosov and G. Gompper}}},\ }\bibfield  {title}
	{\enquote {\bibinfo {title} {Sharp-edged geometric obstacles in microfluidics
				promote deformability-based sorting of cells},}\ }\href@noop {} {\bibfield
		{journal} {\bibinfo  {journal} {Phys. Rev. Fluids}\ }\textbf {\bibinfo
			{volume} {4}},\ \bibinfo {pages} {024201} (\bibinfo {year}
		{2019})}\BibitemShut {NoStop}%
	\bibitem [{\citenamefont {Secomb}(2017)}]{Secomb:2017.1}%
	\BibitemOpen
	\bibfield  {author} {\bibinfo {author} {\bibfnamefont {T.~W.}\ \bibnamefont
			{Secomb}},\ }\bibfield  {title} {\enquote {\bibinfo {title} {{Blood Flow in
					Microcirculation}},}\ }\href@noop {} {\bibfield  {journal} {\bibinfo
			{journal} {Annu. Rev. Fluid Mech.}\ }\textbf {\bibinfo {volume} {49}},\
		\bibinfo {pages} {443} (\bibinfo {year} {2017})}\BibitemShut {NoStop}%
	\bibitem [{\citenamefont {Cantat}\ and\ \citenamefont
		{Misbah}(1999)}]{Misbah:1999.1}%
	\BibitemOpen
	\bibfield  {author} {\bibinfo {author} {\bibfnamefont {I.}~\bibnamefont
			{Cantat}}\ and\ \bibinfo {author} {\bibfnamefont {C.}~\bibnamefont
			{Misbah}},\ }\bibfield  {title} {\enquote {\bibinfo {title} {{Lift Force and
					Dynamical unbinding of Adhering Vesicles under Shear Flow}},}\ }\href@noop {}
	{\bibfield  {journal} {\bibinfo  {journal} {Phys. Rev. Lett.}\ }\textbf
		{\bibinfo {volume} {83}},\ \bibinfo {pages} {880} (\bibinfo {year}
		{1999})}\BibitemShut {NoStop}%
	\bibitem [{\citenamefont {Seifert}(1999)}]{Seifert:1999.1}%
	\BibitemOpen
	\bibfield  {author} {\bibinfo {author} {\bibfnamefont {U.}~\bibnamefont
			{Seifert}},\ }\bibfield  {title} {\enquote {\bibinfo {title} {{Hydrodynamic
					Lift on Bound Vesicles}},}\ }\href@noop {} {\bibfield  {journal} {\bibinfo
			{journal} {Phys. Rev. Lett.}\ }\textbf {\bibinfo {volume} {83}},\ \bibinfo
		{pages} {876} (\bibinfo {year} {1999})}\BibitemShut {NoStop}%
	\bibitem [{\citenamefont {Abkarian}\ \emph {et~al.}(2002)\citenamefont
		{Abkarian}, \citenamefont {Lartigue},\ and\ \citenamefont
		{Viallat}}]{Viallat:2002.1}%
	\BibitemOpen
	\bibfield  {author} {\bibinfo {author} {\bibfnamefont {M.}~\bibnamefont
			{Abkarian}}, \bibinfo {author} {\bibfnamefont {C.}~\bibnamefont {Lartigue}},
		\ and\ \bibinfo {author} {\bibfnamefont {A.}~\bibnamefont {Viallat}},\
	}\bibfield  {title} {\enquote {\bibinfo {title} {{Tank Treading and Unbinding
					of Deformable Vesicles in Shear Flow: Determination of the Lift Force}},}\
	}\href@noop {} {\bibfield  {journal} {\bibinfo  {journal} {Phys. Rev. Lett.}\
		}\textbf {\bibinfo {volume} {88}},\ \bibinfo {pages} {068103} (\bibinfo
		{year} {2002})}\BibitemShut {NoStop}%
	\bibitem [{\citenamefont {Leal}(1980)}]{Leal:1980.1}%
	\BibitemOpen
	\bibfield  {author} {\bibinfo {author} {\bibfnamefont {L.~G.}\ \bibnamefont
			{Leal}},\ }\bibfield  {title} {\enquote {\bibinfo {title} {Particle motions
				in a viscous fluid},}\ }\href@noop {} {\bibfield  {journal} {\bibinfo
			{journal} {Annu. Rev. Fluid Mech.}\ }\textbf {\bibinfo {volume} {12}},\
		\bibinfo {pages} {435} (\bibinfo {year} {1980})}\BibitemShut {NoStop}%
	\bibitem [{\citenamefont {Mandal}\ \emph {et~al.}(2015)\citenamefont {Mandal},
		\citenamefont {Bandopadhyay},\ and\ \citenamefont
		{Chakraborty}}]{Chakraborty:2015.1}%
	\BibitemOpen
	\bibfield  {author} {\bibinfo {author} {\bibfnamefont {S.}~\bibnamefont
			{Mandal}}, \bibinfo {author} {\bibfnamefont {A.}~\bibnamefont
			{Bandopadhyay}}, \ and\ \bibinfo {author} {\bibfnamefont {S.}~\bibnamefont
			{Chakraborty}},\ }\bibfield  {title} {\enquote {\bibinfo {title} {{Effect of
					interfacial slip on the cross-stream migration of a drop in an unbounded
					Poiseuille flow}},}\ }\href@noop {} {\bibfield  {journal} {\bibinfo
			{journal} {Phys. Rev. E}\ }\textbf {\bibinfo {volume} {92}},\ \bibinfo
		{pages} {023002} (\bibinfo {year} {2015})}\BibitemShut {NoStop}%
	\bibitem [{\citenamefont {Kaoui}\ \emph {et~al.}(2008)\citenamefont {Kaoui},
		\citenamefont {Ristow}, \citenamefont {Cantat}, \citenamefont {Misbah},\ and\
		\citenamefont {Zimmermann}}]{Kaoui:2008.1}%
	\BibitemOpen
	\bibfield  {author} {\bibinfo {author} {\bibfnamefont {B.}~\bibnamefont
			{Kaoui}}, \bibinfo {author} {\bibfnamefont {G.~H.}\ \bibnamefont {Ristow}},
		\bibinfo {author} {\bibfnamefont {I.}~\bibnamefont {Cantat}}, \bibinfo
		{author} {\bibfnamefont {C.}~\bibnamefont {Misbah}}, \ and\ \bibinfo {author}
		{\bibfnamefont {W.}~\bibnamefont {Zimmermann}},\ }\bibfield  {title}
	{\enquote {\bibinfo {title} {Lateral migration of a two-dimensional vesicle
				in unbounded poiseuille flow},}\ }\href@noop {} {\bibfield  {journal}
		{\bibinfo  {journal} {Phys. Rev. E}\ }\textbf {\bibinfo {volume} {77}},\
		\bibinfo {pages} {021903} (\bibinfo {year} {2008})}\BibitemShut {NoStop}%
	\bibitem [{\citenamefont {Coupier}\ \emph {et~al.}(2008)\citenamefont
		{Coupier}, \citenamefont {Kaoui}, \citenamefont {Podgorski},\ and\
		\citenamefont {Misbah}}]{Misbah:2008.1}%
	\BibitemOpen
	\bibfield  {author} {\bibinfo {author} {\bibfnamefont {G.}~\bibnamefont
			{Coupier}}, \bibinfo {author} {\bibfnamefont {B.}~\bibnamefont {Kaoui}},
		\bibinfo {author} {\bibfnamefont {T.}~\bibnamefont {Podgorski}}, \ and\
		\bibinfo {author} {\bibfnamefont {C.}~\bibnamefont {Misbah}},\ }\bibfield
	{title} {\enquote {\bibinfo {title} {{Noninertial lateral migration of
					vesicles in bounded Poiseuille flow}},}\ }\href@noop {} {\bibfield  {journal}
		{\bibinfo  {journal} {Phys. Fluids}\ }\textbf {\bibinfo {volume} {20}},\
		\bibinfo {pages} {111702} (\bibinfo {year} {2008})}\BibitemShut {NoStop}%
	\bibitem [{\citenamefont {Doddi}\ and\ \citenamefont
		{Bagchi}(2008)}]{Bagchi:2008.1}%
	\BibitemOpen
	\bibfield  {author} {\bibinfo {author} {\bibfnamefont {S.~K.}\ \bibnamefont
			{Doddi}}\ and\ \bibinfo {author} {\bibfnamefont {P.}~\bibnamefont {Bagchi}},\
	}\bibfield  {title} {\enquote {\bibinfo {title} {Lateral migration of a
				capsule in a plane poiseuille flow in a channel},}\ }\href@noop {} {\bibfield
		{journal} {\bibinfo  {journal} {Int. J. Multiph. Flow}\ }\textbf {\bibinfo
			{volume} {34}},\ \bibinfo {pages} {966} (\bibinfo {year} {2008})}\BibitemShut
	{NoStop}%
	\bibitem [{\citenamefont {Farutin}\ and\ \citenamefont
		{Misbah}(2014)}]{Farutin:2014.1}%
	\BibitemOpen
	\bibfield  {author} {\bibinfo {author} {\bibfnamefont {A.}~\bibnamefont
			{Farutin}}\ and\ \bibinfo {author} {\bibfnamefont {C.}~\bibnamefont
			{Misbah}},\ }\bibfield  {title} {\enquote {\bibinfo {title} {{Symmetry
					breaking and cross-streamline migration of three-dimensional vesicles in an
					axial Poiseuille flow}},}\ }\href@noop {} {\bibfield  {journal} {\bibinfo
			{journal} {Phys. Rev. E}\ }\textbf {\bibinfo {volume} {89}},\ \bibinfo
		{pages} {042709} (\bibinfo {year} {2014})}\BibitemShut {NoStop}%
	\bibitem [{\citenamefont {F\"ortsch}\ \emph {et~al.}(2017)\citenamefont
		{F\"ortsch}, \citenamefont {Laumann}, \citenamefont {Kienle},\ and\
		\citenamefont {Zimmermann}}]{Laumann:2017.2}%
	\BibitemOpen
	\bibfield  {author} {\bibinfo {author} {\bibfnamefont {A.}~\bibnamefont
			{F\"ortsch}}, \bibinfo {author} {\bibfnamefont {M.}~\bibnamefont {Laumann}},
		\bibinfo {author} {\bibfnamefont {D.}~\bibnamefont {Kienle}}, \ and\ \bibinfo
		{author} {\bibfnamefont {W.}~\bibnamefont {Zimmermann}},\ }\bibfield  {title}
	{\enquote {\bibinfo {title} {Migration reversal of soft particles in vertical
				flows},}\ }\href@noop {} {\bibfield  {journal} {\bibinfo  {journal} {EPL}\
		}\textbf {\bibinfo {volume} {119}},\ \bibinfo {pages} {64003} (\bibinfo
		{year} {2017})}\BibitemShut {NoStop}%
	\bibitem [{\citenamefont {Laumann}\ \emph
		{et~al.}(2019{\natexlab{a}})\citenamefont {Laumann}, \citenamefont {Schmidt},
		\citenamefont {Farutin}, \citenamefont {Kienle}, \citenamefont {F\"orster},
		\citenamefont {Misbah},\ and\ \citenamefont {Zimmermann}}]{Laumann:2019.1}%
	\BibitemOpen
	\bibfield  {author} {\bibinfo {author} {\bibfnamefont {M.}~\bibnamefont
			{Laumann}}, \bibinfo {author} {\bibfnamefont {W.}~\bibnamefont {Schmidt}},
		\bibinfo {author} {\bibfnamefont {A.}~\bibnamefont {Farutin}}, \bibinfo
		{author} {\bibfnamefont {D.}~\bibnamefont {Kienle}}, \bibinfo {author}
		{\bibfnamefont {S.}~\bibnamefont {F\"orster}}, \bibinfo {author}
		{\bibfnamefont {C.}~\bibnamefont {Misbah}}, \ and\ \bibinfo {author}
		{\bibfnamefont {W.}~\bibnamefont {Zimmermann}},\ }\bibfield  {title}
	{\enquote {\bibinfo {title} {{Emerging Attractor in Wavy Poiseuille Flows
					Triggers Sorting of Biological Cells}},}\ }\href@noop {} {\bibfield
		{journal} {\bibinfo  {journal} {Phys. Rev. Lett.}\ }\textbf {\bibinfo
			{volume} {122}},\ \bibinfo {pages} {128002} (\bibinfo {year}
		{2019}{\natexlab{a}})}\BibitemShut {NoStop}%
	\bibitem [{\citenamefont {Dincau}\ \emph {et~al.}(2020)\citenamefont {Dincau},
		\citenamefont {Dressaire},\ and\ \citenamefont {Sauret}}]{Sauret:2020.1}%
	\BibitemOpen
	\bibfield  {author} {\bibinfo {author} {\bibfnamefont {B.}~\bibnamefont
			{Dincau}}, \bibinfo {author} {\bibfnamefont {E.}~\bibnamefont {Dressaire}}, \
		and\ \bibinfo {author} {\bibfnamefont {A.}~\bibnamefont {Sauret}},\
	}\bibfield  {title} {\enquote {\bibinfo {title} {{Pulsatile Flow in
					Microfluidic Systems}},}\ }\href@noop {} {\bibfield  {journal} {\bibinfo
			{journal} {Small}\ }\textbf {\bibinfo {volume} {16}},\ \bibinfo {pages}
		{1904032} (\bibinfo {year} {2020})}\BibitemShut {NoStop}%
	\bibitem [{\citenamefont {Lafzi}\ \emph {et~al.}(2020)\citenamefont {Lafzi},
		\citenamefont {Raffiee},\ and\ \citenamefont {Dabiri}}]{Daibri:2020.1}%
	\BibitemOpen
	\bibfield  {author} {\bibinfo {author} {\bibfnamefont {A.}~\bibnamefont
			{Lafzi}}, \bibinfo {author} {\bibfnamefont {A.~H.}\ \bibnamefont {Raffiee}},
		\ and\ \bibinfo {author} {\bibfnamefont {S.}~\bibnamefont {Dabiri}},\
	}\bibfield  {title} {\enquote {\bibinfo {title} {{Inertial migration of a
					deformable capsule in an oscillatory flow in a microchannel}},}\ }\href@noop
	{} {\bibfield  {journal} {\bibinfo  {journal} {Phys. Rev. E}\ }\textbf
		{\bibinfo {volume} {102}},\ \bibinfo {pages} {063110} (\bibinfo {year}
		{2020})}\BibitemShut {NoStop}%
	\bibitem [{\citenamefont {Recktenwald}\ \emph {et~al.}(2021)\citenamefont
		{Recktenwald}, \citenamefont {Wagner},\ and\ \citenamefont
		{John}}]{JohnT:2021.1}%
	\BibitemOpen
	\bibfield  {author} {\bibinfo {author} {\bibfnamefont {S.~M.}\ \bibnamefont
			{Recktenwald}}, \bibinfo {author} {\bibfnamefont {C.}~\bibnamefont {Wagner}},
		\ and\ \bibinfo {author} {\bibfnamefont {T.}~\bibnamefont {John}},\
	}\bibfield  {title} {\enquote {\bibinfo {title} {Optimizing pressure-driven
				pulsatile flows in microfluidic devices},}\ }\href@noop {} {\bibfield
		{journal} {\bibinfo  {journal} {Lab Chip}\ }\textbf {\bibinfo {volume}
			{21}},\ \bibinfo {pages} {2605} (\bibinfo {year} {2021})}\BibitemShut
	{NoStop}%
	\bibitem [{\citenamefont {Loutherback}\ \emph {et~al.}(2009)\citenamefont
		{Loutherback}, \citenamefont {Puchalla}, \citenamefont {Austin},\ and\
		\citenamefont {Sturm}}]{Austin:2009.1}%
	\BibitemOpen
	\bibfield  {author} {\bibinfo {author} {\bibfnamefont {K.}~\bibnamefont
			{Loutherback}}, \bibinfo {author} {\bibfnamefont {J.}~\bibnamefont
			{Puchalla}}, \bibinfo {author} {\bibfnamefont {R.~H.}\ \bibnamefont
			{Austin}}, \ and\ \bibinfo {author} {\bibfnamefont {J.~C.}\ \bibnamefont
			{Sturm}},\ }\bibfield  {title} {\enquote {\bibinfo {title} {{Deterministic
					Microfluidic Ratchet}},}\ }\href@noop {} {\bibfield  {journal} {\bibinfo
			{journal} {Phys. Rev. Lett.}\ }\textbf {\bibinfo {volume} {102}},\ \bibinfo
		{pages} {045301} (\bibinfo {year} {2009})}\BibitemShut {NoStop}%
	\bibitem [{\citenamefont {McFaul}\ \emph {et~al.}(2012)\citenamefont {McFaul},
		\citenamefont {Lin},\ and\ \citenamefont {Ma}}]{McFaul:2012.1}%
	\BibitemOpen
	\bibfield  {author} {\bibinfo {author} {\bibfnamefont {S.~M.}\ \bibnamefont
			{McFaul}}, \bibinfo {author} {\bibfnamefont {B.~K.}\ \bibnamefont {Lin}}, \
		and\ \bibinfo {author} {\bibfnamefont {H.}~\bibnamefont {Ma}},\ }\bibfield
	{title} {\enquote {\bibinfo {title} {{Cell separation based on size and
					deformability using microfluidic funnel ratchets}},}\ }\href@noop {}
	{\bibfield  {journal} {\bibinfo  {journal} {Lab Chip}\ }\textbf {\bibinfo
			{volume} {12}},\ \bibinfo {pages} {2369} (\bibinfo {year}
		{2012})}\BibitemShut {NoStop}%
	\bibitem [{\citenamefont {{E.~S. Park \textit{et al.}}}(2016)}]{ParkE:2016.1}%
	\BibitemOpen
	\bibfield  {author} {\bibinfo {author} {\bibnamefont {{E.~S. Park \textit{et
						al.}}}},\ }\bibfield  {title} {\enquote {\bibinfo {title} {{Continuous Flow
					Deformability-Based Separation of Circulating Tumor Cells Using Microfluidic
					Ratchets}},}\ }\href@noop {} {\bibfield  {journal} {\bibinfo  {journal}
			{Small}\ }\textbf {\bibinfo {volume} {12}},\ \bibinfo {pages} {1909}
		(\bibinfo {year} {2016})}\BibitemShut {NoStop}%
	\bibitem [{\citenamefont {Jo}\ \emph {et~al.}(2016)\citenamefont {Jo},
		\citenamefont {Huang}, \citenamefont {Zimmermann},\ and\ \citenamefont
		{Kanso}}]{Kanso:2016.1}%
	\BibitemOpen
	\bibfield  {author} {\bibinfo {author} {\bibfnamefont {I.}~\bibnamefont
			{Jo}}, \bibinfo {author} {\bibfnamefont {Y.}~\bibnamefont {Huang}}, \bibinfo
		{author} {\bibfnamefont {W.}~\bibnamefont {Zimmermann}}, \ and\ \bibinfo
		{author} {\bibfnamefont {E.}~\bibnamefont {Kanso}},\ }\bibfield  {title}
	{\enquote {\bibinfo {title} {Passive swimming in visous oscillatory flows},}\
	}\href@noop {} {\bibfield  {journal} {\bibinfo  {journal} {Phys. Rev. E}\
		}\textbf {\bibinfo {volume} {94}},\ \bibinfo {pages} {063116} (\bibinfo
		{year} {2016})}\BibitemShut {NoStop}%
	\bibitem [{\citenamefont {Laumann}\ \emph {et~al.}(2017)\citenamefont
		{Laumann}, \citenamefont {Bauknecht}, \citenamefont {Gekle}, \citenamefont
		{Kienle},\ and\ \citenamefont {Zimmermann}}]{Laumann:2017.1}%
	\BibitemOpen
	\bibfield  {author} {\bibinfo {author} {\bibfnamefont {M.}~\bibnamefont
			{Laumann}}, \bibinfo {author} {\bibfnamefont {P.}~\bibnamefont {Bauknecht}},
		\bibinfo {author} {\bibfnamefont {S.}~\bibnamefont {Gekle}}, \bibinfo
		{author} {\bibfnamefont {D.}~\bibnamefont {Kienle}}, \ and\ \bibinfo {author}
		{\bibfnamefont {W.}~\bibnamefont {Zimmermann}},\ }\bibfield  {title}
	{\enquote {\bibinfo {title} {Cross-stream migration of asymmetric particles
				driven by oscillating shear},}\ }\href@noop {} {\bibfield  {journal}
		{\bibinfo  {journal} {EPL}\ }\textbf {\bibinfo {volume} {117}},\ \bibinfo
		{pages} {44001} (\bibinfo {year} {2017})}\BibitemShut {NoStop}%
	\bibitem [{\citenamefont {Morita}\ \emph {et~al.}(2018)\citenamefont {Morita},
		\citenamefont {Omori},\ and\ \citenamefont {Ishikawa}}]{Ishikawa:2018.1}%
	\BibitemOpen
	\bibfield  {author} {\bibinfo {author} {\bibfnamefont {T.}~\bibnamefont
			{Morita}}, \bibinfo {author} {\bibfnamefont {T.}~\bibnamefont {Omori}}, \
		and\ \bibinfo {author} {\bibfnamefont {T.}~\bibnamefont {Ishikawa}},\
	}\bibfield  {title} {\enquote {\bibinfo {title} {{Passive swimming of a
					microcapsule in vertical fluid oscillation}},}\ }\href@noop {} {\bibfield
		{journal} {\bibinfo  {journal} {Phys. Rev. E}\ }\textbf {\bibinfo {volume}
			{98}},\ \bibinfo {pages} {023108} (\bibinfo {year} {2018})}\BibitemShut
	{NoStop}%
	\bibitem [{\citenamefont {{B.~R. Mutlu, J.~F. Edd, and M.
				Toner}}(2018)}]{Mutlu7682}%
	\BibitemOpen
	\bibfield  {author} {\bibinfo {author} {\bibnamefont {{B.~R. Mutlu, J.~F.
					Edd, and M. Toner}}},\ }\bibfield  {title} {\enquote {\bibinfo {title}
			{{Oscillatory inertial focusing in infinite microchannels}},}\ }\href@noop {}
	{\bibfield  {journal} {\bibinfo  {journal} {Proc. Natl. Acad. Sci. U.S.A.}\
		}\textbf {\bibinfo {volume} {115}},\ \bibinfo {pages} {7682} (\bibinfo {year}
		{2018})}\BibitemShut {NoStop}%
	\bibitem [{\citenamefont {Laumann}\ \emph
		{et~al.}(2019{\natexlab{b}})\citenamefont {Laumann}, \citenamefont
		{F\"ortsch}, \citenamefont {Kanso},\ and\ \citenamefont
		{Zimmermann}}]{Laumann:2019.2}%
	\BibitemOpen
	\bibfield  {author} {\bibinfo {author} {\bibfnamefont {M.}~\bibnamefont
			{Laumann}}, \bibinfo {author} {\bibfnamefont {A.}~\bibnamefont {F\"ortsch}},
		\bibinfo {author} {\bibfnamefont {E.}~\bibnamefont {Kanso}}, \ and\ \bibinfo
		{author} {\bibfnamefont {W.}~\bibnamefont {Zimmermann}},\ }\bibfield  {title}
	{\enquote {\bibinfo {title} {Engineering microswimmers by shaking liquids},}\
	}\href@noop {} {\bibfield  {journal} {\bibinfo  {journal} {New J. Phys.}\
		}\textbf {\bibinfo {volume} {21}},\ \bibinfo {pages} {073012} (\bibinfo
		{year} {2019}{\natexlab{b}})}\BibitemShut {NoStop}%
	\bibitem [{\citenamefont {{J. Hendricks, T. Kawakatsu, K. Kawasaki, and W.
				Zimmermann}}(1995)}]{PhysRevE.51.2658}%
	\BibitemOpen
	\bibfield  {author} {\bibinfo {author} {\bibnamefont {{J. Hendricks, T.
					Kawakatsu, K. Kawasaki, and W. Zimmermann}}},\ }\bibfield  {title} {\enquote
		{\bibinfo {title} {{Confined semiflexible polymer chains}},}\ }\href@noop {}
	{\bibfield  {journal} {\bibinfo  {journal} {Phys. Rev. E}\ }\textbf {\bibinfo
			{volume} {51}},\ \bibinfo {pages} {2658} (\bibinfo {year}
		{1995})}\BibitemShut {NoStop}%
	\bibitem [{\citenamefont {Barth\`es-Biesel}(2016)}]{BarthesBiesel:2016.1}%
	\BibitemOpen
	\bibfield  {author} {\bibinfo {author} {\bibfnamefont {D.}~\bibnamefont
			{Barth\`es-Biesel}},\ }\bibfield  {title} {\enquote {\bibinfo {title}
			{{Motion and Deformation of Elastic Capsules and Vesicles in Flow}},}\
	}\href@noop {} {\bibfield  {journal} {\bibinfo  {journal} {Annu. Rev. Fluid
				Mech.}\ }\textbf {\bibinfo {volume} {48}},\ \bibinfo {pages} {25} (\bibinfo
		{year} {2016})}\BibitemShut {NoStop}%
	\bibitem [{\citenamefont {{G. Gompper}}\ and\ \citenamefont {{D.M.
				Kroll}}(1996)}]{Gompper:1996.1}%
	\BibitemOpen
	\bibfield  {author} {\bibinfo {author} {\bibnamefont {{G. Gompper}}}\ and\
		\bibinfo {author} {\bibnamefont {{D.M. Kroll}}},\ }\bibfield  {title}
	{\enquote {\bibinfo {title} {Random surface discretizations and the
				renormalization of the bending rigidity},}\ }\href@noop {} {\bibfield
		{journal} {\bibinfo  {journal} {J. Phys. I France}\ }\textbf {\bibinfo
			{volume} {6}},\ \bibinfo {pages} {1305} (\bibinfo {year} {1996})}\BibitemShut
	{NoStop}%
	\bibitem [{\citenamefont {Kr{\"u}ger}\ \emph {et~al.}(2013)\citenamefont
		{Kr{\"u}ger}, \citenamefont {Gross}, \citenamefont {Raabe},\ and\
		\citenamefont {Varnik}}]{KruegerT:2013.1}%
	\BibitemOpen
	\bibfield  {author} {\bibinfo {author} {\bibfnamefont {T.}~\bibnamefont
			{Kr{\"u}ger}}, \bibinfo {author} {\bibfnamefont {M.}~\bibnamefont {Gross}},
		\bibinfo {author} {\bibfnamefont {D.}~\bibnamefont {Raabe}}, \ and\ \bibinfo
		{author} {\bibfnamefont {F.}~\bibnamefont {Varnik}},\ }\bibfield  {title}
	{\enquote {\bibinfo {title} {Crossover from tumbling to tank-treading-like
				motion in dense simulated suspensions of red blood cells},}\ }\href@noop {}
	{\bibfield  {journal} {\bibinfo  {journal} {Soft Matter}\ }\textbf {\bibinfo
			{volume} {9}},\ \bibinfo {pages} {9008} (\bibinfo {year} {2013})}\BibitemShut
	{NoStop}%
	\bibitem [{\citenamefont {Dhont}(1996)}]{Dhont:96}%
	\BibitemOpen
	\bibfield  {author} {\bibinfo {author} {\bibfnamefont {J.~K.~G.}\
			\bibnamefont {Dhont}},\ }\href@noop {} {\emph {\bibinfo {title} {An
				Introduction to dynamics of colloids}}}\ (\bibinfo  {publisher} {Elsevier},\
	\bibinfo {address} {Amsterdam},\ \bibinfo {year} {1996})\BibitemShut
	{NoStop}%
	\bibitem [{\citenamefont {{J. Rotne and S. Prager}}(1969)}]{Rotne:1969.1}%
	\BibitemOpen
	\bibfield  {author} {\bibinfo {author} {\bibnamefont {{J. Rotne and S.
					Prager}}},\ }\bibfield  {title} {\enquote {\bibinfo {title} {{Variational
					Treatment of Hydrodynamic Interaction in Polymers}},}\ }\href@noop {}
	{\bibfield  {journal} {\bibinfo  {journal} {J. Chem. Phys.}\ }\textbf
		{\bibinfo {volume} {50}},\ \bibinfo {pages} {4831} (\bibinfo {year}
		{1969})}\BibitemShut {NoStop}%
	\bibitem [{\citenamefont {{E. Wajnryb, K.~A. Mizerski, P.~J. Zuk and P.
				Szymczak}}(2013)}]{Wajnryb:2013.1}%
	\BibitemOpen
	\bibfield  {author} {\bibinfo {author} {\bibnamefont {{E. Wajnryb, K.~A.
					Mizerski, P.~J. Zuk and P. Szymczak}}},\ }\bibfield  {title} {\enquote
		{\bibinfo {title} {{Generalization of the Rotne-Prager-Yamakawa mobility and
					shear disturbance tensors}},}\ }\href@noop {} {\bibfield  {journal} {\bibinfo
			{journal} {J. Fluid Mech.}\ }\textbf {\bibinfo {volume} {731}},\ \bibinfo
		{pages} {R3} (\bibinfo {year} {2013})}\BibitemShut {NoStop}%
	\bibitem [{\citenamefont {R.~Skalak}\ and\ \citenamefont
		{Chien}(1973)}]{Skalak:1973.1}%
	\BibitemOpen
	\bibfield  {author} {\bibinfo {author} {\bibfnamefont {R.~P.~Zarda}\
			\bibnamefont {R.~Skalak}, \bibfnamefont {A.~Tozeren}}\ and\ \bibinfo {author}
		{\bibfnamefont {S.}~\bibnamefont {Chien}},\ }\bibfield  {title} {\enquote
		{\bibinfo {title} {{Strain} {Energy} {Function} of {Red} {Blood} {Cell}
				{Membranes}},}\ }\href@noop {} {\bibfield  {journal} {\bibinfo  {journal}
			{Biophys. J.}\ }\textbf {\bibinfo {volume} {13}},\ \bibinfo {pages} {245}
		(\bibinfo {year} {1973})}\BibitemShut {NoStop}%
	\bibitem [{\citenamefont {Meyer}\ \emph {et~al.}(2003)\citenamefont {Meyer},
		\citenamefont {Desbrun}, \citenamefont {Schr{\"o}der},\ and\ \citenamefont
		{Barr}}]{MeyerM:2003.1}%
	\BibitemOpen
	\bibfield  {author} {\bibinfo {author} {\bibfnamefont {M.}~\bibnamefont
			{Meyer}}, \bibinfo {author} {\bibfnamefont {M.}~\bibnamefont {Desbrun}},
		\bibinfo {author} {\bibfnamefont {P.}~\bibnamefont {Schr{\"o}der}}, \ and\
		\bibinfo {author} {\bibfnamefont {A.~H.}\ \bibnamefont {Barr}},\ }\bibfield
	{title} {\enquote {\bibinfo {title} {Discrete differential-geometry operators
				for triangulated 2-manifolds},}\ }in\ \href@noop {} {\emph {\bibinfo
			{booktitle} {Visualization and Mathematics III}}},\ \bibinfo {editor} {edited
		by\ \bibinfo {editor} {\bibfnamefont {H.~C.}\ \bibnamefont {Hege}}\ and\
		\bibinfo {editor} {\bibfnamefont {K.}~\bibnamefont {Polthier}}}\ (\bibinfo
	{publisher} {Springer Berlin Heidelberg},\ \bibinfo {address} {Berlin,
		Heidelberg},\ \bibinfo {year} {2003})\ p.~\bibinfo {pages} {35}\BibitemShut
	{NoStop}%
	\bibitem [{\citenamefont {{C.~S. Peskin}}(2002)}]{Peskin:2002.1}%
	\BibitemOpen
	\bibfield  {author} {\bibinfo {author} {\bibnamefont {{C.~S. Peskin}}},\
	}\bibfield  {title} {\enquote {\bibinfo {title} {The immersed boundary
				method},}\ }\href@noop {} {\bibfield  {journal} {\bibinfo  {journal} {Acta
				Numer.}\ }\textbf {\bibinfo {volume} {11}},\ \bibinfo {pages} {479} (\bibinfo
		{year} {2002})}\BibitemShut {NoStop}%
	\bibitem [{\citenamefont {Kr{\"u}ger}\ \emph {et~al.}(2011)\citenamefont
		{Kr{\"u}ger}, \citenamefont {Varnik},\ and\ \citenamefont
		{Raabe}}]{KruegerT:2011.1}%
	\BibitemOpen
	\bibfield  {author} {\bibinfo {author} {\bibfnamefont {T.}~\bibnamefont
			{Kr{\"u}ger}}, \bibinfo {author} {\bibfnamefont {F.}~\bibnamefont {Varnik}},
		\ and\ \bibinfo {author} {\bibfnamefont {D.}~\bibnamefont {Raabe}},\
	}\bibfield  {title} {\enquote {\bibinfo {title} {{Efficient and accurate
					simulations of deformable particles immersed in a fluid using a combined
					immersed boundary lattice Boltzmann finite element method}},}\ }\href@noop {}
	{\bibfield  {journal} {\bibinfo  {journal} {Comp. Math. Appl.}\ }\textbf
		{\bibinfo {volume} {61}},\ \bibinfo {pages} {3485} (\bibinfo {year}
		{2011})}\BibitemShut {NoStop}%
	\bibitem [{\citenamefont {Kr\"uger}\ \emph {et~al.}(2016)\citenamefont
		{Kr\"uger}, \citenamefont {Kusumaatmaja}, \citenamefont {Kuzmin},
		\citenamefont {Shardt}, \citenamefont {Silva},\ and\ \citenamefont
		{Viggen}}]{KruegerT:2016}%
	\BibitemOpen
	\bibfield  {author} {\bibinfo {author} {\bibfnamefont {T.}~\bibnamefont
			{Kr\"uger}}, \bibinfo {author} {\bibfnamefont {H.}~\bibnamefont
			{Kusumaatmaja}}, \bibinfo {author} {\bibfnamefont {A.}~\bibnamefont
			{Kuzmin}}, \bibinfo {author} {\bibfnamefont {O.}~\bibnamefont {Shardt}},
		\bibinfo {author} {\bibfnamefont {G.}~\bibnamefont {Silva}}, \ and\ \bibinfo
		{author} {\bibfnamefont {E.~M.}\ \bibnamefont {Viggen}},\ }\href@noop {}
	{\emph {\bibinfo {title} {The Lattice Boltzmann Method - Principles and
				Practice}}}\ (\bibinfo  {publisher} {Springer},\ \bibinfo {address}
	{Berlin},\ \bibinfo {year} {2016})\BibitemShut {NoStop}%
	\bibitem [{sup()}]{supplement}%
	\BibitemOpen
	\href@noop {} {}\bibinfo {note} {{ See Supplemental Material at [URL will be
			inserted by publisher] for a LBM simulation of multiple RBCs, a detailed
			description of the two simulation methods, the particle models, and the
			analytical calculation for the net progress of the minimal
			model.}}\BibitemShut {Stop}%
	\bibitem [{all()}]{all_parameters}%
	\BibitemOpen
	\href@noop {} {}\bibinfo {note} {{ Parameters for the MM: $\eta = 1$, $u_1 =
			40$, $u_2 = -20$, $a = 0.1$, $b = 2$, $d = 3$, $k = 60$, simulation only:
			Time step $\delta t = 2 \times 10^{-5}$, $T_1 = 0.6$, $T_2 = 1.2$, run time
			$t_\text{end} = 10 (T_1 + T_2)$. Parameters for the ring polymer: $\delta t =
			2 \times 10^{-4}$, $\eta = 1$, $a = 0.1$, $r_0 = 3$, $k = 10$, $\kappa = 50$,
			$w = 6$, $u_1 = 30$, $u_2 = -15$, $T_1 = 10$, $T_2 = 20$, $t_\text{end} = 5
			\times (T_1+T_2)$.\\ Parameters for the capsule: $\delta t = 5 \times
			10^{-2}$, $\eta = 1$, $a = 0.2$, initial edge length of one triangle $b = 1$
			($r_0 = 6.63$), $\kappa_\text{S} = 0.2$, $\kappa_\text{B} = 0.1$,
			$\kappa_\text{V} = 3$, $w = 20$, $u_1 = 1.5$, $u_2 = -0.75$, $T_1 = 1250$,
			$T_2 = 2500$, $t_\text{end} = 10 \times (T_1+T_2)$.\\ Parameters for the RBC:
			lattice constant $\delta x = 1$, $\delta t = 1$, LBM relaxation time $\tau =
			1$, fluid density $\varrho = 1$ ($\eta = 1/6$), $r_0 = 9$, system size in
			$y$-direction $2 w = 47$, system size in $x$- and $z$-direction (periodic
			boundaries) $S_x = S_z = 128$, $\kappa_\text{S} = 6.51879 \times 10^{-4}$,
			$\kappa_\alpha = 6.51879 \times 10^{-2}$, $\kappa_\text{B} = 2.08293 \times
			10^{-4}$, $\kappa_\text{V} = 6.51879 \times 10^{-2}$, $u_1 = 1.6 \times
			10^{-3}$, $u_2 = -4 \times 10^{-4}$, $T_1 = 2 \times 10^5$, $T_2 = 8 \times
			10^5$, $t_\text{end} = 3 \times (T_1+T_2)$ }}\BibitemShut {NoStop}%
	\bibitem [{\citenamefont {Guckenberger}\ \emph {et~al.}(2018)\citenamefont
		{Guckenberger}, \citenamefont {Kihm}, \citenamefont {John}, \citenamefont
		{Wagner},\ and\ \citenamefont {Gekle}}]{Guckenberger:2018.1}%
	\BibitemOpen
	\bibfield  {author} {\bibinfo {author} {\bibfnamefont {A.}~\bibnamefont
			{Guckenberger}}, \bibinfo {author} {\bibfnamefont {A.}~\bibnamefont {Kihm}},
		\bibinfo {author} {\bibfnamefont {T.}~\bibnamefont {John}}, \bibinfo {author}
		{\bibfnamefont {C.}~\bibnamefont {Wagner}}, \ and\ \bibinfo {author}
		{\bibfnamefont {S.}~\bibnamefont {Gekle}},\ }\bibfield  {title} {\enquote
		{\bibinfo {title} {Numerical--experimental observation of shape bistability
				of red blood cells flowing in a microchannel},}\ }\href@noop {} {\bibfield
		{journal} {\bibinfo  {journal} {Soft Matter}\ }\textbf {\bibinfo {volume}
			{14}},\ \bibinfo {pages} {2032} (\bibinfo {year} {2018})}\BibitemShut
	{NoStop}%
	\bibitem [{\citenamefont {{O. Otto \textit{et al.}}}(2015)}]{GuckJ:2015.1}%
	\BibitemOpen
	\bibfield  {author} {\bibinfo {author} {\bibnamefont {{O. Otto \textit{et
						al.}}}},\ }\bibfield  {title} {\enquote {\bibinfo {title} {{Real time
					deformability cytometry: on-the-fly cell mechanical phenotyping}},}\
	}\href@noop {} {\bibfield  {journal} {\bibinfo  {journal} {Nat. Methods}\
		}\textbf {\bibinfo {volume} {12}},\ \bibinfo {pages} {199} (\bibinfo {year}
		{2015})}\BibitemShut {NoStop}%
	\bibitem [{\citenamefont {Mietke}\ \emph {et~al.}(2015)\citenamefont {Mietke},
		\citenamefont {Otto}, \citenamefont {Girardo}, \citenamefont {Rosendahl},
		\citenamefont {Taubenberger}, \citenamefont {Golfier}, \citenamefont
		{Ulbricht}, \citenamefont {Aland}, \citenamefont {Guck},\ and\ \citenamefont
		{Fischer-Friedrich}}]{GuckJ:2015.2}%
	\BibitemOpen
	\bibfield  {author} {\bibinfo {author} {\bibfnamefont {A.}~\bibnamefont
			{Mietke}}, \bibinfo {author} {\bibfnamefont {O.}~\bibnamefont {Otto}},
		\bibinfo {author} {\bibfnamefont {S.}~\bibnamefont {Girardo}}, \bibinfo
		{author} {\bibfnamefont {P.}~\bibnamefont {Rosendahl}}, \bibinfo {author}
		{\bibfnamefont {A.}~\bibnamefont {Taubenberger}}, \bibinfo {author}
		{\bibfnamefont {S.}~\bibnamefont {Golfier}}, \bibinfo {author} {\bibfnamefont
			{E.}~\bibnamefont {Ulbricht}}, \bibinfo {author} {\bibfnamefont
			{S.}~\bibnamefont {Aland}}, \bibinfo {author} {\bibfnamefont
			{J.}~\bibnamefont {Guck}}, \ and\ \bibinfo {author} {\bibfnamefont
			{E.}~\bibnamefont {Fischer-Friedrich}},\ }\bibfield  {title} {\enquote
		{\bibinfo {title} {Real time deformability cytometry: on-the-fly cell
				mechanical phenotyping},}\ }\href@noop {} {\bibfield  {journal} {\bibinfo
			{journal} {Biophys. J.}\ }\textbf {\bibinfo {volume} {109}},\ \bibinfo
		{pages} {2023} (\bibinfo {year} {2015})}\BibitemShut {NoStop}%
	\bibitem [{\citenamefont {Villone}\ \emph {et~al.}(2016)\citenamefont
		{Villone}, \citenamefont {Greco}, \citenamefont {Hulsen},\ and\ \citenamefont
		{Maffettone}}]{Villone:2016.1}%
	\BibitemOpen
	\bibfield  {author} {\bibinfo {author} {\bibfnamefont {M.~M.}\ \bibnamefont
			{Villone}}, \bibinfo {author} {\bibfnamefont {F.}~\bibnamefont {Greco}},
		\bibinfo {author} {\bibfnamefont {M.~A.}\ \bibnamefont {Hulsen}}, \ and\
		\bibinfo {author} {\bibfnamefont {P.~L.}\ \bibnamefont {Maffettone}},\
	}\bibfield  {title} {\enquote {\bibinfo {title} {Numerical simulations of
				deformable particle lateral migration in tube flow of newtonian and
				viscoelastic media},}\ }\href@noop {} {\bibfield  {journal} {\bibinfo
			{journal} {J. Non-Newton. Fluid Mech.}\ }\textbf {\bibinfo {volume} {234}},\
		\bibinfo {pages} {105} (\bibinfo {year} {2016})}\BibitemShut {NoStop}%
	\bibitem [{\citenamefont {{S.~W. Krauss, P.-Y. Gires and M.
				Weiss}}(2021)}]{krauss}%
	\BibitemOpen
	\bibfield  {author} {\bibinfo {author} {\bibnamefont {{S.~W. Krauss, P.-Y.
					Gires and M. Weiss}}},\ }\href@noop {} {\enquote {\bibinfo {title}
			{{Deformation-induced actuation of cells in asymmetric periodic flow
					fields}},}\ }\bibinfo {howpublished} {bioRxiv preprint,
		doi.org/10.1101/2021.09.30.462560} (\bibinfo {year} {2021})\BibitemShut
	{NoStop}%
	\bibitem [{\citenamefont {Suresh}(2006)}]{suresh_2006}%
	\BibitemOpen
	\bibfield  {author} {\bibinfo {author} {\bibfnamefont {S.}~\bibnamefont
			{Suresh}},\ }\bibfield  {title} {\enquote {\bibinfo {title} {Mechanical
				response of human red blood cells in health and disease: Some
				structure-property-function relationships},}\ }\href {\doibase
		10.1557/jmr.2006.0260} {\bibfield  {journal} {\bibinfo  {journal} {J. Mater.
				Res.}\ }\textbf {\bibinfo {volume} {21}},\ \bibinfo {pages} {1871} (\bibinfo
		{year} {2006})}\BibitemShut {NoStop}%
\end{thebibliography}
\end{document}


\title{
	Oscillating, non-progressing flows induce directed cell motion \\
	- Supplementary information
}

\author{Winfried Schmidt}
\affiliation{Theoretische Physik, Universit\"at Bayreuth, 95440 Bayreuth, Germany}
\affiliation{Laboratoire Interdisciplinaire de Physique, Universit\'e Grenoble Alpes and CNRS, F-38000 Grenoble, France}

\author{Andre F\"ortsch}
\affiliation{Theoretische Physik, Universit\"at Bayreuth, 95440 Bayreuth, Germany}

\author{Matthias Laumann}
\affiliation{Theoretische Physik, Universit\"at Bayreuth, 95440 Bayreuth, Germany}

\author{Walter Zimmermann}
\affiliation{Theoretische Physik, Universit\"at Bayreuth, 95440 Bayreuth, Germany}

\date{October 3, 2021}
\maketitle
\onecolumngrid
\setlength{\parindent}{0cm}
\subsection{LBM simulation of multiple red blood cells}
In the following, we apply our propulsion mechanism to suspensions of several particles and show that RBCs can be sorted according to their deformability.
We consider a LBM simulation where RBCs are initially aligned at the channel center at positions $x_0^1 = 0$, $x_0^2 = 31.5$, $x_0^3 = 63.5$, and $x_0^4 = 95.5$, where $x_0^i$ is the $x$-component of the initial position of the $i$-th RBC. We simulate two RBCs with a shear modulus of $\kappa_\text{S} = 6.51879 \times 10^{-4}$ ("soft", RBCs 1 and 3), and two RBCs with $\kappa_\text{S} = 1.30376 \times 10^{-3}$ ("stiff", RBCs 2 and 4). We employ $u_1 = 2.4 \times 10^{-3}$ ($u_2 = -6 \times 10^{-4}$), the remaining parameters are as in the main text.

\begin{figure}[htb]
	\begin{center}
		\includegraphics[width=0.8\columnwidth]{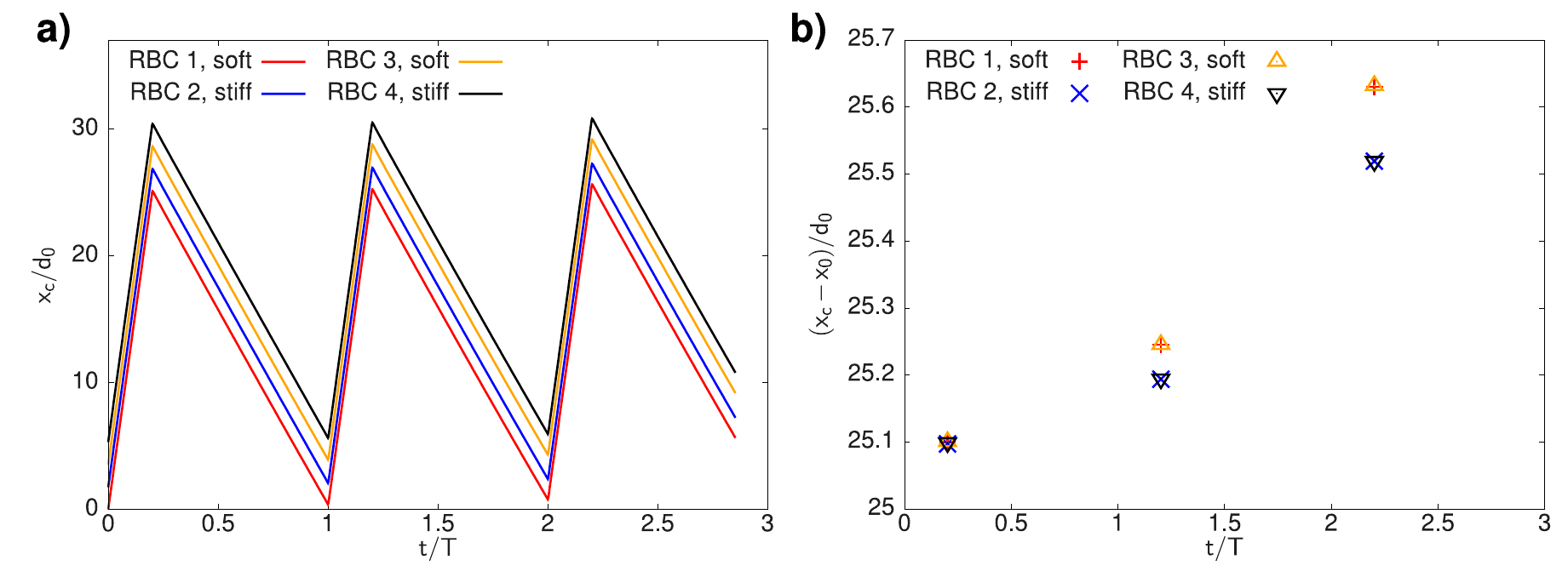}
	\end{center}
	\vspace{-0.4cm}
	\caption{
		LBM simulation of four RBCs with two different elasticities (see legend).
		RBC position [a)] $x_\text{c}(t)$ along the channel axis and respective maxima [b)] in units of their undeformed diameter $d_0$ as function of time $t$. 
	}
	\label{fig_SI_multi_RBCs}
\end{figure}
As shown in Fig.\ \ref{fig_SI_multi_RBCs}a), all RBCs follow the oscillating flow, similar to a single RBC [cf.\ Fig.\ 1c) in the main text]. By displaying the maxima of the trajectories with subtracted initial position in Fig.\ \ref{fig_SI_multi_RBCs}b), it becomes visible that soft RBCs are propelled further than stiff ones. Moreover, the difference in the net progress between two RBCs of the same stiffness is small compared to the difference of the propulsion step between RBCs of different elasticity.
Furthermore, these findings indicate that HI between particles does not play a significant role for separation of RBCs via the asymmetrically oscillating flow.
\subsection{Equations of motion: Simulations in unbounded flows} 
For the MM and the ring polymer we use the Oseen tensor \cite{Dhont:96} in the mobility matrix in Eq.~(4) of the main text,
\begin{equation}
\label{eq_methods_oseen_tensor}
\notag H^{\alpha \beta}_{ij} (\bs{r}_{ij}) = \frac{(1 - \delta_{ij})}{8 \pi \eta r_{ij}} \left( \delta_{\alpha \beta} - \frac{r_{ij}^\alpha r_{ij}^\beta}{r_{ij}^2} \right) + \frac{\delta_{ij}}{6 \pi \eta a} \delta_{\alpha \beta}~,
\end{equation}
with $\alpha, \beta = \{x, y, z\}$, and bead-to-bead vector $\bs{r}_{ij} := \bs{r}_{i} - \bs{r}_{j}$, where $r_{ij} = |\bs{r}_{ij}|$ is the distance between bead $i$ and $j$. For the capsule we use the Rotne-Prager tensor \cite{Dhont:96, Rotne:1969.1, Wajnryb:2013.1}
\begin{align}
\label{eq_methods_rotne_prager_cases}
H^{\alpha \beta}_{ij} (\bs{r}_{ij}) = \begin{cases}
H'^{\alpha \beta}_{ij} (\bs{r}_{ij}) \quad &\text{for} \quad r_{ij} > 2 a \\
H''^{\alpha \beta}_{ij} (\bs{r}_{ij}) \quad &\text{for} \quad r_{ij} \leq 2 a,
\end{cases}
\end{align}
with 
\begin{align}
\label{eq_methods_rotne_prager_no_overlap}
H'^{\alpha \beta}_{ij} (\bs{r}_{ij}) = \frac{1}{8 \pi \eta r_{ij}} \left[ \left( 1 + \frac{2 a^2}{3 r_{ij}^2} \right) \delta_{\alpha \beta} + \left( 1 - \frac{2 a^2}{r_{ij}^2} \right) \frac{r_{ij}^\alpha r_{ij}^\beta}{r_{ij}^2} \right]~,
\end{align}
and a correction for the case of overlapping beads
\begin{align}
\label{eq_methods_rotne_prager_overlap}
\notag H''^{\alpha \beta}_{ij} (\bs{r}_{ij}) = \frac{1}{6 \pi \eta a} \left[ \left( 1 - \frac{9 r_{ij}}{32 a} \right) \delta_{\alpha \beta} + \frac{3 r_{ij}}{32 a} \frac{r_{ij}^\alpha r_{ij}^\beta}{r_{ij}^2}\right]~.
\end{align}
\subsection{Simulations in bounded flows: The Lattice-Boltzmann method}
In the LBM the fluid is discretized by a regular 3D lattice with lattice constant $\delta x$. $f_i (\bs{x}, t)$ is the probability distribution at lattice node $\bs{x}$. We use the D3Q19 scheme, as described for example in Refs.\ \cite{KruegerT:2011.1, KruegerT:2016}, i.e.\ a discrete set of velocities $\bs{c}_i$ (with $i=0,1,...,19$) at each lattice node. With the Bhatnagar-Gross-Krook collision operator the distribution function is evolved according to
\begin{equation}\label{eq_methods_lbe}
f_i (\bs{x} + \bs{c}_i \delta t, t + \delta t) = f_i (\bs{x}, t) - \frac{\delta t}{\tau} \left[ f_i (\bs{x}, t) - f_i^\text{eq} (\bs{x}, t) \right] + F_i \delta t~,
\end{equation}
with time step $\delta t$ and relaxation time $\tau$. The equilibrium distribution is given by
\begin{equation}\label{eq_methods_f_eq}
f_i^\text{eq} = w_i \varrho \left[ 1 + \frac{\bs{c}_i \cdot \bs{u}}{c_\text{s}^2} + \frac{\left( \bs{c}_i \cdot \bs{u} \right)^2}{2 c_\text{s}^4} - \frac{\bs{u} \cdot \bs{u}}{2 c_\text{s}^2} \right]~,
\end{equation}
where $w_i$ are the the lattice weights specified by the D3Q19 scheme and $c_\text{s} = 1/\sqrt{3} \delta x / \delta t$ is the speed of sound. External body force densities $\bs{F}^\text{ext}$ are incorporated in Eq.\ (\ref{eq_methods_lbe}) via
\begin{equation}\label{eq_methods_body_force}
F_i = \left( 1 - \frac{1}{2 \tau} \right) w_i \left( \frac{\bs{c}_i - \bs{u}}{c_\text{s}^2} + \frac{\bs{c}_i \cdot \bs{u}}{c_\text{s}^4} \bs{c}_i \right) \cdot \bs{F}^\text{ext}~.
\end{equation}
The macroscopic fluid density and velocity in Eqs.\ (\ref{eq_methods_f_eq}) and (\ref{eq_methods_body_force}) are obtained via 
\begin{align}\label{eq_methods_lbm_macroscopic_variables}
\varrho (\bs{x}, t) = \sum_{i} f_i \quad \text{and} \quad \bs{u} (\bs{x}, t) = \frac{1}{\varrho} \sum_{i} \bs{c}_i f_i + \frac{\delta t}{2 \varrho} \bs{F}^\text{ext}~,
\end{align}
respectively. The fluid viscosity is obtained via $\eta = \varrho c_\text{s}^2 \delta t \left( \tau - 1/2 \right)$. For the coupling of the particle to the fluid we use the immersed boundary method \cite{Peskin:2002.1, KruegerT:2011.1}. In contrast to our Stokesean dynamics  simulations in unbounded flows, with the LBM the oscillating flow amplitude is controlled indirectly by time-dependent volume forces along the channel axis. In $x$- and $z$-direction we employ periodic boundary conditions.
\subsection{Treatment of the soft particles}
For the ring polymer the total potential $E (\bs{r}) = E_\text{H} + E_\text{B}$ is given by the contribution due to the harmonic springs $E_\text{H}$ with spring stiffness $k$ and equilibrium length $b$. We further employ a bending energy \cite{PhysRevE.51.2658}
\begin{equation}\label{eq_methods_bending_potential_ring}
E_\text{B} = -\frac{\kappa}{2} \ln(1- \cos \alpha_i),
\end{equation}
where $\kappa$ is the bending rigidity and $\alpha_i$ the angle between the connection lines from bead $i$ to its two neighboring beads.

The capsule is discretized by $1280$ triangles (faces) with $N = 642$ beads situated at their edges (nodes).
The total potential of the capsule $E (\bs{r}) = E_\text{V} + E_\text{S} + E_\text{B}$ consists of three contributions: The volume energy \cite{KruegerT:2013.1}
\begin{equation}\label{eq_methods_volume_convervation}
E_\text{V} = \frac{\kappa_\text{V}}{V_0} \left( V - V_0 \right)^2,
\end{equation}
where $\kappa_\text{V}$ is the volume modulus, $V$ the instantaneous volume and $V_0$ the reference volume of the capsule's spherical initial shape. The in-plane strain energy of the membrane is obtained via
\begin{equation}\label{eq_methods_strain_density}
E_\text{S} = \sum_{f} A_f^{(0)} \varepsilon_f^\text{S},
\end{equation}
where $f$ is the face index, $A_f^{(0)}$ the area of the undeformed face and $\varepsilon_f^\text{S}$ the strain energy area density. For the capsule the strain-softening behavior of rubber-like materials is modeled by the Neo-Hookean law \cite{BarthesBiesel:1981.1, BarthesBiesel:2016.1}
\begin{equation}
\varepsilon_f^\text{S} (I_1, I_2) = \frac{\kappa_\text{S}}{2} \left( I_1 - 1 + \frac{1}{I_2 + 1} \right),
\end{equation}
with shear-elastic modulus $\kappa_\text{S}$ and in-plane strain invariants $I_1$ and $I_2$ which are related	to the local membrane deformation tensor. Resistance against bending of the membrane is realized by the contribution \cite{Gompper:1996.1} 
\begin{equation}
E_\text{B} = \frac{\kappa_\text{B}}{2} \sum_{i,j} \left( 1 - \cos \beta_{i,j} \right),
\end{equation}
with bending elasticity $\kappa_\text{B}$ and $\beta_{i,j}$ being the angle formed by the normal vectors of two neighboring faces $i$ and $j$.

The biconcave initial shape of the RBC is obtained by shifting the nodes of the spherical mesh according to
\begin{align}\label{eq_rbc_eq_shape}
x(\varrho) = \pm \frac{r_0}{2} \sqrt{1 - \varrho^2} \left( C_0 + C_2 \varrho^2 + C_4 \varrho^4 \right)
\end{align}
assuming that the rotational symmetry axis of the RBC is along the $x$-axis, with $\varrho := \sqrt{y^2+z^2}/r_0$ \cite{KruegerT:2013.1}. The RBC shape parameters are $C_0 = 0.207$, $C_2 = 2.003$ and $C_4 = - 1.123$ \cite{EvansE:1972.1}.
The total potential of the RBC is given by $E (\bs{r}) = E_\text{V} + E_\text{S} + E_\text{B}$ with $E_\text{V}$ as in Eq.\ (\ref{eq_methods_volume_convervation}). The contribution due to in-plane strain forces is given by Eq.\ (\ref{eq_methods_strain_density}) where we use the Skalak-law \cite{Skalak:1973.1, KruegerT:2013.1}
\begin{equation}
\varepsilon_f^\text{S} (I_1, I_2) = \frac{\kappa_\text{S}}{12} (I_1^2 + 2 I_1 - 2 I_2) + \frac{\kappa_\alpha}{12} I_2^2,
\end{equation}
which describes the strain-hardening elastic behavior of the RBC membrane. Here, $\kappa_\text{S}$ is the RBC strain modulus and $\kappa_\alpha$ the area dilation modulus. The bending contribution to the total potential is given by \cite{Canham:1970.1, helfrich1973elastic}
\begin{equation}
E_\text{B} = 2 \kappa_\text{B} \int \left( H - H_0 \right)^2 \text{d}S,
\end{equation}
where $H(\bs{r}_i)$ and $H_0(\bs{r}_i)$ are the local mean and spontaneous curvature of the membrane at node $i$, and $\kappa_\text{B}$ is the bending modulus. The integral runs over the instantaneous surface $S$ of the RBC. For the implementation of the bending forces acting on each node we employ the algorithm proposed by Meyer et al.\ \cite{MeyerM:2003.1} which was shown to perform well when compared to other methods \cite{Guckenberger:2016.1}. The bending force acting on node $i$ is obtained via
\begin{equation}\label{eq_methods_meyer_bending_force}
\bs{F}^\text{B} (\bs{r}_i) = 2 \kappa_\text{B} \left[\Delta_\text{s} (H (\bs{r}_i) - H_0 (\bs{r}_i)) + 2(H (\bs{r}_i) - H_0 (\bs{r}_i)) \left(H (\bs{r}_i)^2 - K (\bs{r}_i) + H (\bs{r}_i) H_0 (\bs{r}_i) \right) \right] A^i_\text{mixed} \hat{n} (\bs{r}_i)~,
\end{equation}
with the local Gaussian curvature $K(\bs{r}_i)$, the local outwards directed normal vector $\hat{n} (\bs{r}_i)$ and the Laplace-Beltrami operator on the surface $\Delta_\text{s}$. The mixed area $A^i_\text{mixed}$ is a fraction of the total surface area that is assigned to each node and is computed according to Ref.\ \cite{MeyerM:2003.1}. The Laplace-Beltrami operator, when applied to a function $\chi (\bs{r}_i)$, is given by
\begin{equation}\label{eq_methods_laplace_beltrami_calc}
\Delta_\text{s} \chi (\bs{r}_i) \approx \frac{1}{2 A^i_\text{mixed}} \sum_{j(i)} \left(\cot \theta_1^{(ij)} + \cot \theta_2^{(ij)} \right) \left[ \chi (\bs{r}_i) - \chi (\bs{r}_j) \right]~,
\end{equation}
where the sum runs over the one-ring neighborhood of the $i$-th node, with $j$ being the index of nodes which form adjacent faces to node $i$. $\theta_1^{(ij)}$ and $\theta_2^{(ij)}$ are the angles at the vertices opposite to the edge $\langle i,j \rangle$ in the triangles containing node $j$ and $j+1$, respectively, which is illustrated e.g.\ in Ref.\ \cite{Graham:2015.1}. The normal vector in Eq.\ (\ref{eq_methods_meyer_bending_force}) is computed by averaging the normal vectors of the adjacent faces, weighted with their incident angle to node $i$ \cite{JinS:2005.1}. The local mean curvature is calculated via
\begin{equation}\label{eq_methods_mean_curvature}
H (\bs{r}_i)= \frac{1}{2} \sum_{k=1}^{3} (\Delta_\text{s} r_{i,k}) \hat{n}_k (\bs{r}_i)~,
\end{equation}
where $\hat{n}_k (\bs{r}_i)$ is the $k$-th component of normal vector at node $i$. The spontaneous curvature is set to zero. Finally, the Gaussian curvature in Eq.\ (\ref{eq_methods_meyer_bending_force}) is obtained via
\begin{equation}\label{eq_methods_gaussian_curvature}
K (\bs {r}_i)=\frac{1}{A^i_\text{mixed}} \left( 2 \pi - \sum_t \theta^{(i)}_t \right)~,
\end{equation}
with the sum running over all triangles adjacent to node $i$ and $\theta^{(i)}_t$ being the angle in triangle $t$ at node $i$.

\subsection{Derivation of the propulsion velocity for the minimal model}\label{sec_appendix}
\begin{figure}[htb]
	\begin{center}
		\includegraphics[width=0.8\columnwidth]{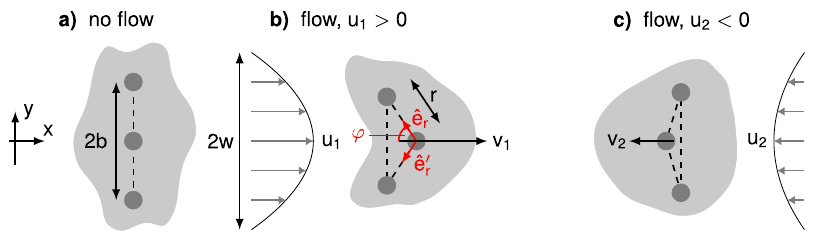}
	\end{center}
	\vspace{-0.4cm}
	\caption{
		a): Undeformed MM in a quiescent fluid. The elastic particle is modeled by three beads connected by three springs (dashed lines) of equilibrium length $b$ and $2b$, respectively. 
		b): Sketch of the MM in the faster forward section of the oscillatory flow with amplitude $u_1 > 0$ and resulting particle velocity $v_1 >0$. The deformation is described by bead-to-bead distance $r$ and opening angle $\varphi$, given by the unit vectors $\hat{e}_r$ and $\hat{e}_r^\prime$ (red).
		c): Shape of the MM in the slower backward flow section with $u_2 <0$ and particle speed $v_2 <0$. The larger flow gradients in the forward section ($u_1 > |u_2|$) cause a larger particle deformation.
	}
	\label{fig_SI_sketch}
\end{figure}

%
Due to the simplicity of the MM an analytical approach for the calculation of the propulsion velocity is possible.
Our assumption of $t_R \ll T_{1,2}$ implies that the particle adapts its respective stationary shape [as sketched in Fig.~(\ref{fig_SI_sketch}) b) and c)] quickly after each change of the flow direction.
With this we can assume that the three beads have the same velocity in each of the two flow sections.
We further neglect particle-wall interactions as well as HI between the beads. The equations of motion thus simplify to 
\begin{equation}\label{eq_of_motion_MM}
	\dot{\bs{r}}_i = \bs{u} (\bs{r}_i) + \frac{\bs{f}_i}{\zeta}
\end{equation}
for bead $i = \{1,2,3 \}$.
The equilibrium positions of the beads are given by $\bs{r}_1 = (0,0)$, $\bs{r}_2 = (0,b)$ and $\bs{r}_3 = (0,-b)$. We further quantify the deformation of the MM by the angle $\varphi$ which is formed by the connection of the center and top bead with the horizontal, and the two unit vectors $\hat{e}_r = \left(  - \cos \varphi, \sin \varphi \right)$ and $\hat{e}_r' = \left(  - \cos \varphi, - \sin \varphi \right)$, as shown in Fig.~(\ref{fig_SI_sketch}) b). Furthermore we introduce the bead-to-bead distance $r = |\bs{r}_1 - \bs{r}_2| = |\bs{r}_1 - \bs{r}_3|$.
The forces acting on each of the three beads can then be expressed as
\begin{align}\label{eq_analytic_forces}
\notag\bs{f}_1 &= k (r-b) \hat{e}_r + k (r-b) \hat{e}'_r, \\
\bs{f}_2 &= - k (r-b) \hat{e}_r - k (2 r \sin \varphi - 2 b) \hat{e}_y, \\
\notag\bs{f}_3 &= - k (r - b) \hat{e}_r' + k  (2 r \sin \varphi - 2 b) \hat{e}_y~,
\end{align}
We start by considering only the forward section of the asymmetrically oscillating parabolic flow.
From Eq.~(\ref{eq_of_motion_MM}) and Eqs.~(\ref{eq_analytic_forces}) we obtain
\begin{align}
\label{eq_eq_of_motion_rod_c}
\dot{\bs{r}}_1 &= \begin{pmatrix} v_1 \\ \dot{y} \end{pmatrix} = \begin{pmatrix} u_1 \\ 0 \end{pmatrix} + \frac{k}{\zeta} (r-b) \left[ \begin{pmatrix} - \cos \varphi \\ \sin \varphi \end{pmatrix} + \begin{pmatrix} - \cos \varphi \\ - \sin \varphi \end{pmatrix} \right], \\
\label{eq_eq_of_motion_rod_t}
\dot{\bs{r}}_2 &= \begin{pmatrix} v_1 \\ \dot{y} \end{pmatrix} = \begin{pmatrix} u_1 \\ 0 \end{pmatrix} \left( 1 - \frac{r^2 \sin^2 \varphi}{w^2} \right) - \frac{k}{\zeta} (r-b) \begin{pmatrix} - \cos \varphi \\ \sin \varphi \end{pmatrix} - 2 \frac{k}{\zeta} \left( r \sin \varphi - b \right) \begin{pmatrix} 0 \\ 1 \end{pmatrix}.
\end{align}
Due to the problem's symmetry it is sufficient to only consider the equations of motion for beads 1 and 2. From the $y$-component of Eq.\ (\ref{eq_eq_of_motion_rod_c}) follows $\dot{y} = 0$ and together with the $y$-component of Eq.\ (\ref{eq_eq_of_motion_rod_t}) we obtain
\begin{equation}\label{eq_analytic_sin_phi}
\sin \varphi = \frac{2}{3 \tilde{R} - 1}~,
\end{equation}
where we introduced the ratio $\tilde{R} := r/b$. Note that $\tilde{R} = 1$ holds for the non-deformed equilibrium state of the MM. Eliminating $\varphi$ via Eq.\ (\ref{eq_analytic_sin_phi}) in the equations for the $x$-coordinates of Eqs.\ (\ref{eq_eq_of_motion_rod_c}) and (\ref{eq_eq_of_motion_rod_t}) yields
\begin{equation}\label{eq_analytic_elimin_phi}
\frac{\sqrt{1 - \left( \frac{2}{3 \tilde{R} - 1} \right)^2} \left( \tilde{R} - 1 \right)}{\frac{C_1}{6} \left( \frac{2 \tilde{R}}{3 \tilde{R} - 1} \right)^2} = 1,
\end{equation}
where we identified $C_1 = 2 u_1 \zeta b / (k w^2)$. Using the approximation of small deformations, by a series expansion of Eq. (\ref{eq_analytic_elimin_phi}) around $\tilde{R} = 1$ we obtain
\begin{equation}
\frac{6 \sqrt{3}}{C_1} \left( \tilde{R} - 1 \right)^\frac{3}{2} \approx 1~,
\end{equation}
when truncated after the first non-vanishing order. Solving for $\tilde{R}$ yields
\begin{align}\label{eq_analytic_r_tilde}
\tilde{R} (C_1) = \frac{\sqrt[3]{2}}{6} C_1^\frac{2}{3} + 1~.
\end{align}
By substituting Eqs.\ (\ref{eq_analytic_r_tilde}) and (\ref{eq_analytic_sin_phi}) in Eqs.\ (\ref{eq_eq_of_motion_rod_c}) we obtain
\begin{equation}\label{eq_analytic_v1}
v_1 = u_1 - \frac{\sqrt[3]{2}}{3} \frac{k b}{\zeta}  C_1^\frac{2}{3} \sqrt{1 + \frac{2}{\frac{\sqrt[3]{2}}{4} C_1^\frac{2}{3} + 1}}
\end{equation}
for the velocity in the forward flow section. Similar considerations lead to the expression for the velocity in the backward flow section,
\begin{equation}\label{eq_analytic_v2}
v_2 = u_2 + \frac{\sqrt[3]{2}}{3} \frac{k b}{\zeta}  C_2^\frac{2}{3} \sqrt{1 + \frac{2}{\frac{\sqrt[3]{2}}{4} C_2^\frac{2}{3} + 1}}
\end{equation}
with $C_2 = 2 |u_2| \zeta b / (k w^2)$. With Eq.~(6) from the main text follows from Eqs.\ (\ref{eq_analytic_v1}) and (\ref{eq_analytic_v2})
\begin{equation}\label{eq_analytic_actuation_velocity_final_app}
v_\text{p} = \frac{\sqrt[3]{2} C_1^\frac{2}{3} b}{3 \left(A + 1\right) t_R} \left[ A^\frac{1}{3} \sqrt{1 + \frac{8}{\sqrt[3]{2} \left( \frac{C_1}{A} \right)^\frac{2}{3} + 4}} -  \sqrt{1 + \frac{8}{\sqrt[3]{2} C_1^\frac{2}{3} + 4}} \right]
\end{equation}
where we used $u_2 = - u_1/A$ and $C_2 = C_1/A$. With $B_{1,2} := \left(2 C_{1,2}^2\right)^{1/3}$ and $W(X):=\sqrt{1 + 8X^{\frac{2}{3}}    \left(4X^{\frac{2}{3}} +B_1\right)^{-1}}$ Eq.~(\ref{eq_analytic_actuation_velocity_final_app}) can be written as
\begin{align}
v_\text{p} \approx \frac{b}{3 t_R} ~ B_1  \frac{W(A) A^{\frac{1}{3}} - W(1)}{A+1}.
\end{align}
Using Eqs.~(\ref{eq_analytic_sin_phi}) and (\ref{eq_analytic_r_tilde}), the lateral size of the MM during $T_1$ and $T_2$ is given by
\begin{equation}
	\Delta y_{1,2} = 2 r_{1,2} \sin \varphi_{1,2} = \frac{4b \left(\frac{B_{1,2}}{6}+1\right)}{\left(\frac{B_{1,2}}{2}+2\right)} \approx 2b \left[1- \frac{B_{1,2}}{12} +  \mathcal{O} (B_{1,2}^2)\right].
\end{equation}